\documentclass[11pt]{article}
\usepackage{amsmath,amsfonts}
\usepackage{todonotes}
\usepackage{geometry}
\usepackage{caption}
\usepackage{cite}
\usepackage{setspace}
\usepackage{url}
\usepackage{comment}
\usepackage{hyperref}
\usepackage{rotating}
\newcommand{\be}{\begin{equation}}
\newcommand{\ee}{\end{equation}}

\newcommand{\ON}[1]{\mathrm{O}( #1 )}
\newcommand{\SU}[1]{\mathrm{SU}( #1 )}
\newcommand{\SL}[1]{\mathrm{SL}( #1 )}

\newcommand{\SO}[1]{\mathrm{SO}( #1 )}
\newcommand{\EG}[1]{\mathrm{E}_{#1(#1)}}

\newcommand{\Spin}[1]{\mathrm{Spin}(#1)}

\newcommand{\USp}[1]{\mathrm{USp}(#1)}

\newcommand{\hpartial}{\hat{\partial}}

\newcommand{\bfo}{\bar{4}}
\newcommand{\bfi}{\bar{5}}
 
\newcommand{\bi}{\bar{i}} 
\newcommand{\bj}{\bar{j}} 
\newcommand{\bk}{\bar{k}} 
\newcommand{\bl}{\bar{l}}
\newcommand{\bm}{\bar{m}}

\newcommand{\bp}{\bar{p}}

\newcommand{\obf}[1]{\overline{\mathbf{#1}}}
\newcommand{\mbf}[1]{\mathbf{#1}}

\newcommand{\gL}{\mathcal{L}}

\newcommand{\gM}{\mathcal{M}}

\newcommand{\x}{x^1}
\newcommand{\y}{x^2}
\newcommand{\z}{x^3}
\newcommand{\w}{x^4}
\newcommand{\xf}{x^5}
\newcommand{\xsi}{x^6}
\newcommand{\xse}{x^7}
\newcommand{\dx}{d\x}
\newcommand{\dy}{d\y}
\newcommand{\dz}{d\z}
\newcommand{\dw}{d\w}
\newcommand{\dxf}{d\xf}
\newcommand{\dxsi}{d\xsi}
\newcommand{\dxse}{d\xse}
\newcommand{\pdx}{\left(\dx\right)}
\newcommand{\pdy}{\left(\dy\right)}
\newcommand{\pdz}{\left(\dz\right)}
\newcommand{\pdw}{\left(\dw\right)}
\newcommand{\pdxf}{\left(\dxf\right)}
\newcommand{\pdxsi}{\left(\dxsi\right)}
\newcommand{\pdxse}{\left(\dxse\right)}
\numberwithin{equation}{section}

\newcommand\Tstrut{\rule{0pt}{3ex}}         
\newcommand\Bstrut{\rule[-1.3ex]{0pt}{0pt}}   

\begin{document}

\begin{titlepage}
\vfill

\begin{flushright}
LMU--ASC 63/17 \\
MPP--2017--223
\end{flushright}

\vfill

\begin{center}
   \baselineskip=16pt
   	{\LARGE \bf Locally non-geometric fluxes and \\ \vskip0.2cm missing momenta in M-theory}
   	\vskip 2cm
 	{\large \bf Dieter L\"ust$^{a,b}$\footnote{dieter.luest@lmu.de}, Emanuel Malek$^a$\footnote{e.malek@lmu.de}, Marc Syv\"ari$^a$\footnote{marc.syvaeri@physik.uni-muenchen.de}}
   	\vskip .6cm
   	{\it $^a$ Arnold Sommerfeld Center for Theoretical Physics, Department f\"ur Physik, \\ Ludwig-Maximilians-Universit\"at M\"unchen, Theresienstra{\ss}e 37, 80333 M\"unchen, Germany\\ \ \\
   	$^b$ Max-Planck-Institut f\"ur Physik,
   	  Werner-Heisenberg-Institut\\ F\"ohringer Ring 6, 80805 M\"unchen, Germany \\ \ \\}
   	\vskip 1cm
\end{center}
\vfill

\begin{abstract}
We use exceptional field theory to describe locally non-geometric spaces of M-theory of more than three dimensions. For these spaces, we find a new set of locally non-geometric fluxes which lie in the R-R sector in the weak-coupling limit and can typically be characterised by mixed symmetry tensors. These spaces thus provide new examples of non-geometric backgrounds which go beyond the NS-NS sector of string theory. Starting from twisted tori we construct duality chains that lead to these new non-geometric backgrounds and we show that, just as in the four-dimensional case, there are missing momenta associated to the mixed symmetry tensors.
\end{abstract}

\vfill

\end{titlepage}

\tableofcontents

\newpage

\section{Introduction} \label{s:Intro}
One of the characteristic features of string theory is the existence of dualities. For example, because strings can wind around non-trivial cycles, circles of radius $R$ and of radius $\alpha'/R$ are equivalent in string theory, leading to the concept of T-duality. These dualities also suggest that the string target space can be a non-geometric background which is patched together with dualities \cite{Dabholkar:2002sy,Hellerman:2002ax,Kachru:2002sk,Flournoy:2004vn,Hull:2004in,Dabholkar:2005ve}. A subclass of such spaces, called globally non-geometric, still admit a local description in terms of a metric, NS-NS 2-form and possibly R-R fluxes. However, because of the duality patching these fields are not well-defined over the entire background. More generally, we can consider a locally non-geometric background which does not even admit a local description in terms of a metric, NS-NS 2-form and R-R fluxes \cite{Shelton:2005cf}. Non-geometric compactifications could have interesting phenomenological consequences, for example in moduli stabilisation \cite{Flournoy:2004vn}.

Neither class of non-geometric spaces are manifolds and therefore our usual notions of spacetime break down. For example, it was shown in \cite{Lust:2010iy,Blumenhagen:2010hj,Blumenhagen:2011ph,Condeescu:2012sp,Andriot:2012vb,Bakas:2015gia} that globally non-geometric backgrounds are non-commutative, while locally non-geometric spaces also exhibit non-associativity. The simplest kind of locally non-geometric spaces are characterised by a fully antisymmetric spacetime tensor $R^{ijk}$, called the $R$-flux, \cite{Shelton:2005cf} and which gives rise to a closed string phase space algebra \cite{Lust:2010iy,Blumenhagen:2010hj,Blumenhagen:2011ph,Mylonas:2012pg,Bakas:2013jwa}
\begin{equation}
 \left[ x^i,\, x^j \right] = \frac{i \ell_s^3}{\hbar} R^{ijk} p_k \,, \qquad \left[ x^i,\, p_j \right] = i\, \hbar\, \delta^i_j \,, \qquad \big[ p_i,\, p_j \big] = 0 \,.
\end{equation}
The Jacobiator amongst coordinates does not vanish
\begin{equation}
 \left[ x^i,\, x^j,\, x^k \right] = \frac13 \left( \left[ \left[ x^i,\, x^j \right],\, x^k \right] + \left[ \left[ x^k,\, x^i \right],\, x^j \right] + \left[ \left[ x^j,\, x^k \right],\, x^i \right] \right) = \ell_s^3\, R^{ijk} \,, \label{eq:RFluxAlgebra}
\end{equation}
so that the coordinate space becomes non-commutative. This algebra was also be deformation quantised in \cite{Mylonas:2012pg,Mylonas:2013jha} using a membrane model.

Recently, four-dimensional non-geometric backgrounds of M-theory were studied in \cite{Blair:2014zba,Gunaydin:2016axc,Kupriyanov:2017oob,Lust:2017bgx}, particularly focussing on local non-geometry. The lift of the string theory $R$-flux $R^{ijk}$ to a four-dimensional M-theory background was constructed in \cite{Blair:2014zba}, using the duality manifest formulation of exceptional field theory \cite{Berman:2010is}. The four-dimensional M-theory $R$-flux is given by a mixed symmetry tensor $R^{i,jklm}$ with $R^{i,jklm} = R^{i,[jklm]}$ and $R^{[i,jklm]} = 0$.

In \cite{Gunaydin:2016axc}, a toy model of a locally non-geometric M-theory background was considered which is U-dual to a Heisenberg Nilmanifold, or ``twisted torus''. The homology of this twisted torus implies that M2-branes cannot wrap a certain 2-cycle. By U-duality this implies that there is a missing momentum mode in the locally non-geometric background \cite{Gunaydin:2016axc}. As a result the phase space of M2-branes in this four-dimensional locally non-geometric M-theory background is seven- instead of eight-dimensional. In \cite{Gunaydin:2016axc} it was conjectured that in more complicated four-dimensional locally non-geometric backgrounds, the missing momenta are given by
\begin{equation}
 R^{i,jklm} p_i = 0 \,. \label{eq:Missing4dMomentum}
\end{equation}

If we reduce to IIA string theory, the $R^{i,jklm}$ flux reduces to the usual $R^{ijk}$ flux of string theory \cite{Shelton:2005cf}. On the other hand, the missing momentum mode is in that case along the M-theory circle and therefore \eqref{eq:Missing4dMomentum} implies that there are no D0-branes in a string $R$-flux background. As argued in \cite{Wecht:2007wu} there should be no D0-branes in an $R$-flux background because after applying three T-dualities these D0-branes become D3-branes wrapping a 3-cycle with non-trivial $H$-flux. However, such a configuration is forbidden by the Freed-Witten anomaly condition. Furthermore, the absence of D0-branes, and hence point particles, is natural in a non-associative spacetime where one has a ``minimal volume'' \cite{Mylonas:2013jha}.

This fact that there is a missing momentum mode was crucial in the conjecture of \cite{Gunaydin:2016axc} that the phase space of M2-branes in a locally non-geometric M-theory background is governed by the non-associative algebra of the imaginary octonions, which was subsequently quantised in \cite{Kupriyanov:2017oob}. The reduction to type IIA string theory can then be understood as a contraction of the algebra of imaginary octonions which yields \eqref{eq:RFluxAlgebra}.

In this paper we will further study locally non-geometric backgrounds in M-theory, by using exceptional field theory to find the $R$-fluxes in higher dimensions\footnote{Throughout this paper we will be referring to the dimension of the non-geometric background, not that of the external space.}, and study the relationship between the existence of these fluxes and missing momentum modes. Following \cite{Andriot:2012wx} and \cite{Blair:2014zba} these can be constructed as spacetime tensors involving dual derivatives, which can be thought of as Fourier dual to wrapping modes of branes, of the trivector $\Omega^{ijk}$ or six-vector $\Omega^{ijklmn}$ appearing in the globally well-defined non-geometric parameterisation of the string background. We find a variety of new locally non-geometric $R$-fluxes, most of which have mixed symmetry. For example, in seven dimensions (the case of $\EG{7}$) these are given by
\begin{equation}
 \begin{split}
  R^{i,jklm} &= 4 \hpartial^{i[j} \Omega^{klm]} - e_{\bi}{}^{[j} \epsilon^{klm]inpq} \hpartial_{pq} e^{\bi}{}_n \,, \\
  R^{ij}{}_k &= \hpartial_{kl} \Omega^{ijl} - \frac1{72} \epsilon_{klmnpqr} \hpartial^{ij} \Omega^{lmnpqr} - \frac{1}{36} \epsilon_{klmnpqr} \Omega^{lmn} \hpartial^{ij} \Omega^{pqr} \\
  & \quad + 4 e_{\bi}{}^{[i} \hpartial^{j]} e^{\bi}{}_k + 4 \delta^{[i}_k e_{\bi}{}^{j]} \hpartial^l e^{\bi}{}_l \,, \\
  R^i &= \hpartial_{jk} \Omega^{ijk} - 4 e_{\bi}{}^j \hpartial^i e^{\bi}{}_j - 8 e_{\bi}{}^i \hpartial^j e^{\bi}{}_j \,, \\
  R &= \epsilon_{ijklmnp} \hpartial^{i} \Omega^{jklmnp} - 2 \epsilon_{ijklmnp} \Omega^{ijk} \hpartial^m \Omega^{lnp} \,, \\
  R^{ijkl} &= \frac58 \hpartial^{[i} \Omega^{jkl]} + \frac12 \hpartial_{pq} \Omega^{pqijkl} + \frac14 \Omega^{[ijk} \hpartial_{pq} \Omega^{l]pq} \,,
 \end{split}
\end{equation}
where $\hpartial^{ij}$, $\hpartial_{ij} = \frac1{5!} \epsilon_{ijklmnp} \hpartial^{klmnp}$ and $\hpartial^i = \frac1{7!} \epsilon_{jklmnpq} \hpartial^{i,jkmnpq}$ are improved dual derivatives in seven dimensions which we define in detail in \eqref{eq:E7HatDerivatives}.

We then construct toy models for these locally non-geometric backgrounds by successively applying U-dualities to ${\cal N}_3 \times T^n$ or ${\cal N}_2 \times T^n$ where ${\cal N}_3$ is the Heisenberg Nilmanifold and ${\cal N}_2$ is a two-dimensional non-compact space which has non-unimodular geometric flux and has not been considered in the literature before. Using these duality chains we can deduce from the topology of the initial background that for each mixed symmetry $R$-flux tensor there are missing momentum modes satisfying
\begin{equation}
 \begin{split}
  R^{i,jklm}\, p_i &= 0 \,, \\
  \left( R^{ij}{}_k - 2 R^{[i} \delta^{j]}_k \right) p_i &= 0 \,, \\
  R^i\, p_i &= 0 \,, \\
  R^{ijkl}\, p_i &= 0 \,,
 \end{split}
\end{equation}
providing a nice higher-dimensional generalisation of the conjecture of \cite{Gunaydin:2016axc}. In the case of non-vanishing seven-dimensional singlet flux we find that all momenta vanish, i.e.
\begin{equation}
 R\, p_i = 0 \,.
\end{equation}
Just as in \cite{Gunaydin:2016axc}, the conditions $R^{i,jklm} p_i = 0$, $R^{ijkl} p_i = 0$ and $R\, p_i = 0$ can equivalently be understood as coming from the Freed-Witten anomaly in IIB.

An interesting aspect of this work is that we are using exceptional field theory to describe parallelisable spaces which are not topologically tori and therefore have different homology groups. Often the extra coordinates of EFT are assumed to be related to wrapping modes of branes but this would imply that in the cases considered here the EFT should have fewer coordinates. However, we would argue that there should instead be a different interpretation of the extra coordinates not related to wrapping modes. For example, in double field theory they can be understood as the zero modes of the independent left- and right-movers of the string, as was also argued in DFT on group manifolds \cite{Blumenhagen:2014gva}, and we would hope that there can be a similar interpretation for the EFT coordinates. Indeed, the results obtained here seem consistent with such an interpretation. For example, as mentioned previously, in the four-dimensional case the missing momentum mode can also be understood via the Freed-Witten anomaly \cite{Gunaydin:2016axc,Wecht:2007wu}. Furthermore, its IIA limit is consistent with the fact that there should be no point particles in a non-associative spacetime \cite{Mylonas:2013jha}.

This paper is organised as follows. In section \ref{s:T3Chain} we set the scene by reviewing the prototypical example of a locally non-geometric space in string theory by dualising a three-torus with $H$-flux. Then we use exceptional field theory to construct locally non-geometric fluxes of M-theory backgrounds in higher dimensions in \ref{s:RFlux}. Finally, in section \ref{s:DualityChain} we dualise appropriate geometric backgrounds to construct examples of new locally non-geometric spaces which are characterised by the novel $R$-fluxes. We use the topology of the initial geometric space to deduce which momenta vanish in the locally non-geometric spaces. Finally, we conclude in section \ref{s:Conclusions} with a discussion about the implications for non-associativity and open questions.

\section{Review of string theory $T^3$ duality chain} \label{s:T3Chain}

We begin with a brief recap of the chain of dualities that generates a locally non-geometric string background starting with a $T^3$ with $H$-flux \cite{Kachru:2002sk,Shelton:2005cf}.

\paragraph{$T^3$ with $H$-flux:} Consider a $T^3$ with $H$-flux, $H_{123} = N$. We can write the metric and $B$-field as
\begin{equation}
 \begin{split}
  ds^2 &= \pdx^2 + \pdy^2 + \pdz^2 \,, \\
  B_{12} &= N \z \,.
 \end{split}
\end{equation}

\paragraph{Nilmanifold:} If we perform a T-duality along the $\x$ direction we obtain a twisted torus, with metric
\begin{equation}
 ds^2 = \left( \dx - N \z \dy \right)^2 + \pdy^2 + \pdz^2 \,,
\end{equation}
and no $B$-field. We say this space has geometric flux, because it is parallelisable with well-defined 1-forms
\begin{equation}
 \begin{split}
  e^{\bar{1}} &= \dx - N \z \dx \,, \\
  e^{\bar{2}} &= \dy \,, \\
  e^{\bar{3}} &= \dz \,,
 \end{split}
\end{equation}
and the ``geometric flux'' is given by
\begin{equation}
 de^{\bar{i}} = T_{\bar{j}\bar{k}}{}^{\bar{i}} e^{\bar{j}} \wedge e^{\bar{k}} \,,
\end{equation}
hence $T^1{}_{23} = N$. We see that the coordinates are identified as
\begin{equation}
 \left( \x,\y,\z \right) \sim \left( \x + 1,\y,\z \right) \sim \left( \x,\y + 1,\z \right) \sim \left( \x + N \y,\y,\z + 1 \right) \,,
\end{equation}
and as we discuss in more detail in \ref{s:TwistedTorus}, this does not have the topology of a $T^3$.

This change in topology can be understood as follows \cite{Bouwknegt:2003vb,Bouwknegt:2003wp}. The $T^3$ with $H$-flux can be viewed as a trivial circle bundle over $T^2$. It is useful to define the two-form obtained by integrating the 3-form field strength over the $S^1$ fibre, parameterised by the $\x$ coordinate,
\begin{equation}
 \bar{H} = \int_{S^1} H = N \dy \wedge \dz \in H^2\left(T^2\right) \,.
\end{equation}
After T-duality, we obtain another $S^1$-bundle over $T^2$, where the 1st Chern class of the new $T^2$ bundle is given by the $\bar{H}$, while the new $H$-flux is the old 1st Chern class. Because we initially had the trivial circle bundle $T^3$, we do not obtain a $H$-flux, but the initial $H$-flux leads to a non-trivial $S^1$-bundle with 1st Chern class $c_1 = \bar{H} \in H^2\left(T^2\right)$. This is the Heisenberg Nilmanifold.

\paragraph{Globally non-geometric space:} The way we have written the Nilmanifold, the metric is only independent of $\x$ and $\y$. A duality along $\x$ returns us to the $T^3$ with $H$-flux, and so to continue we must T-dualise along $\y$. However, $\partial_2$ is not a well-defined vector field on the nilmanifold and therefore $\y$ is not the coordinate along a well-defined circle in the nilmanifold. Thus the T-dual is a globally non-geometric space. This can also be seen directly from the Buscher rules which give the following metric and $B$-field,
\begin{equation}
 \begin{split}
  ds^2 &= \frac{\pdx^2 + \pdy^2}{1+N^2\left(\z\right)^2} + \pdz^2 \,,\\
  B_{23} &= \frac{N\z}{1+N^2\left(\z\right)^2} \,, \label{eq:QFlux}
 \end{split}
\end{equation}
neither of which are well-defined as $\z \longrightarrow \z + 1$.

Nonetheless, there is a different, ``non-geometric'', parameterisation of the background in terms of open-string variables
\begin{equation}
 \begin{split}
  \beta^{ij} &= \frac12 \left( \left( g - B \right)^{-1} - \left( g+B \right)^{-1} \right) \,, \\
  \hat{g} &= \frac12 \left( \left(g - B\right)^{-1} + \left(g + B\right)^{-1} \right)^{-1} \,.
 \end{split}
\end{equation}
In generalised geometry \cite{Grana:2008yw} and double field theory \cite{Hull:2009mi,Andriot:2011uh,Andriot:2012an} these new variables correspond to a different parameterisation of the generalised vielbein, which encodes the metric, and which is related to the usual one by a local $\ON{d} \times \ON{d}$ rotation.

For the space \eqref{eq:QFlux}, the metric and bivector $\beta$ are then given by
\begin{equation}
 \hat{ds}^2 = \pdx^2 + \pdy^2 + \pdz^2 \,, \qquad \beta^{12} = N\z \,, \label{eq:QFluxBeta}
\end{equation}
with $\hat{ds}^2$ the line element of $\hat{g}$. This generalised vielbein in this parameterisation is globally well-defined, while the local $\ON{d} \times \ON{d}$ rotation relating \eqref{eq:QFluxBeta} and \eqref{eq:QFlux} is not. As a result, it is the ``non-geometric parameterisation'' in terms of $\hat{g}$ and $\beta$ which correctly describes the global structure of this non-geometric space. This background is classified by its ``$Q$-flux'' which is a spacetime tensor \cite{Andriot:2012wx,Andriot:2013xca} and is in this case given by
\begin{equation}
 Q_i{}^{jk} = \partial_i \beta^{jk} + \ldots \,,
\end{equation}
with the terms denotes by $\ldots$ necessary for covariance but which vanish in this case. For the space \eqref{eq:QFluxBeta} its only non-zero components are
\begin{equation}
 Q_3{}^{12} = N \,.
\end{equation}
In the following we will simplify the notation by dropping the hat on the metric of the non-geometric parameterisation $\hat{g} \longrightarrow g$.

\paragraph{Locally non-geometric space:} The only other duality we can consider is along the $\z$ direction, which is not an isometry due to $\beta^{12} = N \z$ and therefore the Buscher rules cannot be used. However, one should still be able to make sense of T-duality in this case \cite{Dabholkar:2005ve}. We know that the entire spectrum of string excitations of T-dual circles are equivalent. We can now consider a coherent state of string momentum excitations on top of a circle background, which would deform the metric and $B$-field of the circle so that it is no longer the standard round metric, and the isometry is lost. Under T-duality, this configuration becomes a coherent state of string winding excitations on top of the dual circle background.

This coherent state of string winding excitations is captured in the doubled geometry as a background depending on the dual coordinates, as has for example been argued in \cite{Dabholkar:2005ve}. This is exactly what happens after applying a T-duality without isometries in doubled models \cite{Hull:2004in,Hull:2006va} such as double field theory \cite{Hull:2009mi}, where extra coordinates are introduced which are dual to winding modes. Then a T-duality along $\z$ exchanges the coordinate with its T-dual coordinate $\z \longleftrightarrow \tilde{x}_1$, and without isometry the T-dual background will depend on the dual coordinates $\tilde{x}_1$, and is called locally non-geometric.

Applying the T-duality along the $\z$ coordinate of the $Q$-flux background we obtain
\begin{equation}
 \begin{split}
  \hat{ds}^2 &= \left(dx^2\right)^2 + \left(dx^2\right)^2 + \left(dx^3\right)^2 \,, \\
  \beta^{12} &= N \tilde{x}_3 \,. \label{eq:RFluxT3}
 \end{split}
\end{equation}
The local non-geometry is measured by a spacetime tensor known as the $R$-flux \cite{Andriot:2011uh,Andriot:2012wx}, which is given by
\begin{equation}
 R^{ijk} = 3 \hat{\partial}^{[i} \beta^{jk]} \,,
\end{equation}
with $\hat{\partial}^{i} = \tilde{\partial}^{i} + \beta^{ij} \partial_j$. For \eqref{eq:RFluxT3} one finds
\begin{equation}
 R^{123} = N \,.
\end{equation}

We see that T-duality just shifts the value of $N$, given by the initial $H$-flux to the geometric and non-geometric fluxes. This is often summarised by the diagram
\begin{equation}
 H_{123} \stackrel{T_{1}}{\longrightarrow} T_{23}{}^1
\stackrel{T_{2}}{\longrightarrow} Q_{3}{}^{12}
\stackrel{T_{3}}{\longrightarrow} R^{123} \,,
\end{equation}
where $T_i$, with $i = 1, \ldots, 3$, denotes a T-duality along $x^i$. The ``non-geometric fluxes'' can also arise when applying a duality along a timelike direction, where, at least formally, the same structures $Q_i{}^{jk}$ and $R^{ijk}$ can appear \cite{Malek:2012pw,Malek:2013sp}.

In this paper we will be interested in generalising this chain to M-theory to generate and study examples of new locally non-geometric backgrounds which we discuss next in section \ref{s:RFlux}. We then generalise the above duality chain in section \ref{s:DualityChain}.

\section{Locally non-geometric fluxes in M-theory} \label{s:RFlux}
We now wish to find local expressions for the locally non-geometric fluxes which appear more than four dimensions. Here we follow \cite{Andriot:2012wx,Andriot:2012an,Blair:2014zba}, and use a ``non-geometric parameterisation'' of the exceptional generalised vielbein in terms of a metric and trivector \cite{Malek:2012pw} and the higher-dimensional generalisation. This parameterisation is in turn globally well-defined on the non-geometric background and allows us to find local expressions, known as the $R$-flux, characterising the locally non-geometric backgrounds. As shown in \cite{Andriot:2012wx}, these fluxes are spacetime tensors under diffeomorphisms, and we will use this criterion to find their local expressions. We note that for the NS-NS sector of type II string theory, the non-geometric fluxes can also be understood using Lie algebroids as in \cite{Blumenhagen:2012nk,Blumenhagen:2012nt,Blumenhagen:2013aia}.

Recently, the non-geometric parameterisation of the exceptional generalised vielbein was used in \cite{Lee:2016qwn} to derive parts of higher-dimensional globally non-geometric $Q$-fluxes, which in \cite{Lee:2016qwn} were called $S$-fluxes, under certain simplifying assumptions. Here we will complete these expressions for the $Q$-fluxes of \cite{Lee:2016qwn} so that they are spacetime tensors.

We begin with a brief review of the essential features of exceptional field theory \cite{Berman:2010is} in subsection \ref{s:EFT} followed by a review of the four-dimensional case \cite{Blair:2014zba} in subsection \ref{s:SL5RFlux} to explain our methodology. We then generalise the treatment to larger dimensions in subsections \ref{s:SO55RFlux} -- \ref{s:E7RFlux}.

\subsection{Relevant features of $\EG{d}$ EFT} \label{s:EFT}
In this paper we will follow \cite{Blair:2014zba} and use exceptional field theory (EFT) \cite{Berman:2010is,Berman:2011cg,Berman:2011pe,Berman:2012vc,Hohm:2013pua,Aldazabal:2013via,Hohm:2013vpa,Hohm:2013uia} to describe higher-dimensional locally non-geometric backgrounds in M-theory. We begin by reviewing the essential features of EFT, which provides a U-duality manifest reformulation of compactifications of M-theory to $D$ dimensions. The continuous version of U-dualities acting on a $d = 11-D$-dimensional toroidal compactification form the split real forms of the exceptional groups, which we have listed together with their maximal compact subgroups $\mathrm{H}_d$ in table \ref{t:ExcGroups}.

\vskip1em
\noindent\makebox[\textwidth]{
 \begin{minipage}{\textwidth}
  \begin{center}
  \begin{tabular}{|c|c|c|c|}
  \hline
  $D$ & $\EG{d}$ & ${\mathrm H}_d$ & $R_1$\Tstrut\Bstrut \\ \hline
  7 & $\SL{5}$ & $\USp{4}$ & $\mbf{10}$ \\
  6 & $\Spin{5,5}$ & $\USp{4}\times\USp{4}$ & $\mbf{16}$ \\
  5 & $\EG{6}$ & $\USp{8}$ & $\mbf{27}$ \\
  4 & $\EG{7}$ & $\SU{8}$ & $\mbf{56}$ \\
   \hline
  \end{tabular}
  \vskip-0.5em
  \captionof{table}{\small{Split real forms of the exceptional groups, their maximal compact subgroups (ignoring discrete factors) and their $R_1$ representation.}}\label{t:ExcGroups} 
  \end{center}
 \end{minipage}}
\vspace{1em}

In EFT and the related exceptional generalised geometry \cite{Grana:2009im,Coimbra:2011ky,Coimbra:2012af}, the bosonic (fermionic) degrees of freedom of 11-dimensional supergravity at each point are combined into representations of $\EG{d}$ ($\mathrm{H}_d$). For example, the fully internal bosonic degrees of freedom (metric $g_{ij}$, 3-form $C_{ijk}$ and 6-form $C_{ijklmn}$) are combined into the generalised metric $\gM$ which parameterises the coset space
\begin{equation}
 \gM \in \frac{\EG{d}}{\mathrm{H}_d} \,.
\end{equation}
Therefore, it can be written in terms of generalised vielbeine $E^{\bar{M}}{}_M$, which are $\EG{d}$ group elements, as
\begin{equation}
 \gM_{MN} = E^{\bar{M}}{}_M E^{\bar{N}}{}_N \delta_{\bar{M}\bar{N}} \,,
\end{equation}
where $M$, $N = 1, \ldots, \mathrm{dim}\,R_1$ label the $R_1$ representation of $\EG{d}$ listed in table \ref{t:ExcGroups}. The $\mathrm{H}_d$ subgroup of $\EG{d}$ acts as local transformations on the ``flat indices'' $\bar{M}$, $\bar{N}$ and preserves the inner product $\delta_{\bar{M}\bar{N}}$.

Similarly, the generators of the local symmetries, vector fields (for diffeomorphisms), 2-forms and 5-forms (for gauge transformations) also combine into the $R_1$ representation. Furthermore, all bosonic degrees of freedom with 1 leg on the external space also transform in the $R_1$ representation, while those with $n > 1$ legs on the external legs transform under other representations, typically called $R_n$ \cite{Cederwall:2013naa,Aldazabal:2013via,Hohm:2015xna,Wang:2015hca}, which also appear in the tensor hierarchy of maximal gauged supergravities \cite{deWit:2008ta,deWit:2008gc,Kleinschmidt:2011vu,Palmkvist:2011vz}.

In exceptional generalised geometry these fields are sections of generalised tangent bundles ${\cal R}_1 \simeq TM \oplus \Lambda^2 T^*M \oplus \Lambda^5 T^*M \oplus \ldots$ where $M$ is the usual physical spacetime. The exceptional groups are then the structure group of these vector bundles \cite{Coimbra:2011ky}. However, having a fixed physical background $M$ breaks U-duality covariance. Instead, EFT makes use of ``extended spaces'' $\tilde{M}$ with coordinates $X^M$ transforming in the dual of the $R_1$ representation. On a torus these coordinates are Fourier dual to momentum and possible wrapping modes of branes, and therefore transform under U-dualities. The local symmetries (diffeomorphisms and gauge transformations) are realised as a local $\EG{d}$ action on the fields generated by the generalised Lie derivative \cite{Coimbra:2011ky,Berman:2011cg,Coimbra:2012af,Berman:2012vc}, for example acting on a vector field in the $R_1$ representation, $V \in \Gamma\left({\cal R}_1\right)$, as
\begin{equation}
 \gL_{\Lambda} V^M = \Lambda^N \partial_N V^M - V^N \partial_N \Lambda^M + Y^{MP}_{NQ} V^N \partial_P \Lambda^Q \,.
\end{equation}
Here $\Lambda^M \in \Gamma\left({\cal R}_1\right)$, consisting of vector field, 2-form and 5-form, is the generator of the ``generalised diffeomorphism'', $\partial_M = \partial / \partial X^M$ and $Y^{MP}_{NQ}$ is an invariant tensor of $\EG{d}$ \cite{Berman:2012vc}. We give explicit expressions for these invariants in appendix \ref{A:EFT}. In what follows, we will read off the transformation properties of spacetime fields from the generalised Lie derivative.

For the algebra of generalised diffeomorphisms to close we impose the ``section condition''
\begin{equation} 
 Y^{MN}_{PQ} \partial_M \otimes \partial_N = 0 \,,
\end{equation}
when the derivatives act on any pair of fields or as double derivatives of any field. This can be solved by the dependence of all fields on $d$ coordinates, corresponding to the physical internal spacetime. An alternative solution restrict the dependence to $d-1$ coordinates, corresponding to an internal spacetime of IIB string theory \cite{Blair:2013gqa,Hohm:2013vpa}.

The generalised metric $\gM \in \EG{d} / \mathrm{H}_d$ can be parameterised by ``geometric'' fields $g_{ij}$, $C_{ijk}$, $C_{ijklmn}$ or by ``non-geometric ones'' $g_{ij}$, $\Omega^{ijk}$, $\Omega^{ijklmn}$. The latter is called a non-geometric parameterisation because it is this parameterisation that is globally well-defined on a non-geometric background, just like for $g_{ij}$ and $\beta^{ij}$ \cite{Grana:2008yw,Andriot:2011uh,Andriot:2012an} in the non-geometric torus example discussed in section \ref{s:T3Chain}. The generalised Lie derivative determines the way these different fields transform under spacetime diffeomorphisms and gauge transformations.

Finally, let us mention that if we consider a truncation of EFT on a generalised parallelisable background \cite{Lee:2014mla,Berman:2013uda,Hohm:2014qga,duBosque:2017dfc,Inverso:2017lrz}, so that the generalised vector fields $E_{\bar{M}}{}^M$ are globally well-defined, then this reduction will yield a $D$-dimensional maximal gauged SUGRA with embedding tensor
\begin{equation}
 \theta_{\bar{M}\bar{N}}{}^{\bar{P}} = E^{\bar{P}}{}_M \gL_{E_{\bar{M}}} E_{\bar{N}}{}^M \,,
\end{equation}
which is the torsion of the Weitzenb\"ock connection \cite{Coimbra:2011nw,Berman:2013uda,Blair:2014zba}. One can check that an embedding tensor obtained this way automatically satisfies the linear constraint of maximal gauged SUGRAs which determine in what representations of $\EG{d}$ it can lie.

\subsection{Four-dimensional locally non-geometric backgrounds} \label{s:SL5RFlux}
Four-dimensional compactifications are described by the $\SL{5}$ exceptional field theory, where $R_1 = \mbf{10}$, or antisymmetric representation, of $\SL{5}$. Therefore the $\SL{5}$ EFT has 10 ``extended coordinates'' $X^{ab} = \left( X^{i5} ,\, X^{ij} \right) = \left( x^i ,\, \frac12 \epsilon^{ijkl} \tilde{x}_{ij} \right)$, where $a = 1, \ldots, 5$, $i = 1, \ldots, 4$ and $X^{ab} = X^{[ab]}$. Here the $x^i$ are the usual spacetime coordinates on the physical internal space while $\tilde{x}_{ij}$ are the additional ``wrapping coordinates''. The generalised metric parameterises the coset space
\begin{equation}
 \gM \in \frac{\SL{5}}{\SO{5}} \,,
\end{equation}
and thus can be represented by a $5 \times 5$ matrix of unit determinant. We will here use this representation, rather than the $10 \times 10$ one.

We begin with the $\SL{5}$ generalised vielbein in its non-geometric parameterisation
\begin{equation}
 E^{\bar{a}}{}_a = \begin{pmatrix}
  |e|^{-2/5} e^{\bar{i}}{}_i & 0 \\
  \frac1{3!} |e|^{3/5} \epsilon_{ijkl} \Omega^{jkl} & |e|^{3/5}
 \end{pmatrix} \,, \label{eq:SL5GenVielbein}
\end{equation}
where $\epsilon_{ijkl}$ is the four-dimensional Levi-Civita symbol such that $\epsilon_{1234} = 1$. This differs from the parameterisation given in \cite{Malek:2012pw,Blair:2014zba} by a conformal transformation so that the generalised vielbein here has unit determinant. The determinant factors $|e|$ are required so that $e^{\bar{i}}{}_i$ transforms as a vielbein under spacetime diffeomorphisms. On a generic background, a local $\SO{5}$ transformation can be used to turn this into an upper-triangular matrix, corresponding to a ``geometric parameterisation'' in terms of a metric $g_{ij}$ and a local 3-form $C_{ijk}$. However, on a non-geometric background, the corresponding $\SO{5}$ transformation fails to be well-defined and hence also the ``geometric parameterisation'' $g_{ij}$, $C_{ijk}$ fail to be well-defined. In this paper, we will restrict ourselves to examples where the background is generalised parallelisable, meaning that the inverse generalised vielbein defines a set of globally well-defined generalised vector fields. On such backgrounds, it will be straightforward to see which parameterisation is globally well-defined.

We now act with the generalised Lie derivatives on the generalised vielbein \eqref{eq:SL5GenVielbein}, $\gL_V E^{\bar{a}}{}_a$, to find the transformation law for $\Omega^{ijk}$ under spacetime diffeomorphisms. To do this, we take
\begin{equation}
 V^{ab} = \left(V^{i5} ,\, V^{ij} \right) = \left( \xi^i,\, 0 \right) \,,
\end{equation}
so that only the spacetime diffeomorphism generator is non-zero. This way we find
\begin{equation}
 \begin{split}
  \delta_\xi \Omega^{ijk} &= \xi^l \partial_l \Omega^{ijk} - 3 \Omega^{l[ij} \partial_l \xi^{k]} - 3 \partial^{[ij} \xi^{k]} \,, \\
  &= L_\xi \Omega^{ijjk} - 3 \partial^{[ij} \xi^{k]} \,, \label{eq:SL5OmegaTransform}
 \end{split}
\end{equation}
where $L_\xi$ denotes the action of the usual spacetime Lie derivative on a tensor. The metric $g_{ij} = e^{\bi}{}_i e_{\bi\,j}$ is a spacetime tensor as usual
\begin{equation}
 \delta_\xi g_{ij} = L_\xi g_{ij} = \xi^k \partial_k g_{ij} + 2 g_{k(i} \partial_{j)} \xi^k \,.
\end{equation}

Following the philosophy first introduced in \cite{Andriot:2012wx} to describe the NS-NS sector of type II string theory, one can construct various spacetime tensors out of $g_{ij}$ and $\Omega^{ijk}$, but the one that measures the local non-geometric must involve dual derivatives of the trivector $\Omega^{ijk}$. To construct this, it is worthwhile to first introduce an improved dual derivative $\hpartial^{ij}$ such that when it acts on a scalar field one obtains a spacetime tensor, i.e.
\begin{equation}
 \delta_\xi \hpartial^{ij} \phi = L_\xi \hpartial^{ij} \phi \,,
\end{equation}
up to the section condition
\begin{equation}
 \partial_i \otimes \partial^{ij} + \partial^{ij} \otimes \partial_i = \partial_i \partial^{ij} = 0 \,, \label{eq:SL5SectionCondition}
\end{equation}
where $\otimes$ represents the action of the derivatives on two different fields. Using \eqref{eq:SL5OmegaTransform} one finds that
\begin{equation}
 \hpartial^{ij} = \partial^{ij} + \Omega^{ijk} \partial_k \,.
\end{equation}
This can alternatively be defined as
\begin{equation}
 \hpartial^{ij} = \frac12 |e|^{-4/5} e_{\bi}{}^i e_{\bj}{}^j \epsilon^{\bi\bj\bk\bl}  E_{\bk}{}^a E_{\bl}{}^b \partial_{ab} \,.
\end{equation}

Using these ingredients it is easy to construct a spacetime tensor involving a single dual derivative $\partial^{ij}$ of $\Omega^{ijk}$. The only possibility is given by \cite{Blair:2014zba}
\begin{equation}
 R^{i,jklm} = 4 \hpartial^{i[j} \Omega^{klm]} \,. \label{eq:SL5RFlux}
\end{equation}
The existence of this flux can also be seen from the embedding tensor of maximal gauged supergravities, which encodes all possible gaugings of those theories, and transforms in the $\obf{15} \oplus \obf{40} \oplus \obf{10}$ representations of $\SL{5}$ \cite{Samtleben:2005bp}. Decomposing this under $\SL{4} \times \mathbb{R}^+ \subset \SL{5}$ one finds the following representations
\begin{equation}
 \begin{split}
  \obf{15} &\longrightarrow \obf{10}_2 \oplus \obf{4}_{-3} \oplus \mbf{1}_{-8} \,, \\
  \obf{40} &\longrightarrow \obf{20}_{-3} \oplus \mbf{10}_2 \oplus \mbf{6}_2 \oplus \obf{4}_7 \,, \\
  \obf{10} &\longrightarrow \obf{4}_{-3} \oplus \mbf{6}_2 \,.
 \end{split} \label{eq:SL5EmbeddingTensorSplit}
\end{equation}
Loosely speaking, the higher the $\mathbb{R}^+$ charge, the ``more non-geometric'' the origin of the flux \cite{Samtleben:2008pe}. This statement can be made very precise as follows. From the decomposition of the $\obf{10}$ of $\SL{5}$ we see that the derivatives transform as
\begin{equation}
 \partial_i \in \obf{4}_{-3} \,, \qquad \partial^{ij} \in \mbf{6}_2 \,.
\end{equation}
We can also decompose the adjoint representation of $\SL{5}$ under $\SL{4} \times \mathbb{R}^+$ to find
\begin{equation}
 \mbf{24} \longrightarrow \mbf{15}_0 \oplus \obf{4}_5 \oplus \mbf{4}_{-5} \oplus \mbf{1}_0 \,.
\end{equation}
However, the adjoint representation of the exceptional group corresponds to the fields appearing in the generalised vielbein. This allows us to identify 
\begin{equation}
 e_{\bi}{}^i \in \mbf{15}_0 \oplus \mbf{1}_0 \,, \qquad C_{ijk} \in \mbf{4}_{-5} \,, \qquad \Omega^{ijk} \in \obf{4}_{5} \,.
\end{equation}
Now we clearly see that the $\mbf{4}_7$ representation of the $\obf{40}$ of $\SL{5}$ clearly corresponds to the locally non-geometric flux \eqref{eq:SL5RFlux}:
\begin{equation}
 R^{i,jklm} \in \mbf{4}_7 \,.
\end{equation}
This can also be checked by an explicit calculation of the embedding tensor in terms of the generalised vielbein, see for example \cite{Blair:2014zba}. Furthermore, since the embedding tensor encodes all the fluxes, and it has only one flux, the $\mbf{4}_7$, of the correct $\mathbb{R}^+$ charge to be locally non-geometric, we can be sure to have found all $R$-fluxes. Finally, when appropriately reducing on a circle, this $R$-flux reduces to the string theory $R$-flux in the NS-NS sector \cite{Blair:2014zba,Gunaydin:2016axc}.

\subsubsection{Other non-geometric fluxes} \label{s:SL5Other}
As shown in \cite{Blair:2014zba} we can also construct other fluxes from dual derivatives of $g_{ij}$ or spacetime derivatives of $\Omega^{ijk}$. To do so, we introduce the following connections for both types of derivatives
\begin{equation}
 \begin{split} 
  \Gamma_{ij}{}^{k} &= e_{\bi}{}^{k} \partial_i e^{\bi}{}_{j} \,, \\
  \hat{\Gamma}^{ij}{}_k{}^l &= e_{\bi}{}^{l} \hpartial^{ij} e^{\bi}{}_{k} \,.
 \end{split}
\end{equation}
Under spacetime diffeomorphisms, these transform as
\begin{equation}
 \begin{split} 
  \delta_{\xi} \Gamma_{ij}{}^{k} &= L_{\xi} \Gamma_{ij}{}^{k} + \partial_i \partial_j \xi^k \,, \\
  \delta_{\xi} \hat{\Gamma}^{ij}{}_k{}^l &= L_{\xi} \hat{\Gamma}^{ij}{}_k{}^l + \hpartial^{ij} \partial_k \xi^l \,.
 \end{split}
\end{equation}
$\Gamma_{ij}{}^k$ is the Weitzenb\"ock connection whose torsion is useful in classifying parallelisable backgrounds. Using these objects we can construct the following spacetime tensors in four-dimensions, which will correspond to non-geometric fluxes. These can be made well-defined on generalised parallelisable backgrounds since they appear in components of the embedding tensor.

\paragraph{$Q$-flux:} This is also called the ``globally non-geometric flux'' and is defined as
\begin{equation}
 Q_i{}^{jkl} = \partial_i\Omega^{jkl} + 3 \hat{\Gamma}^{[jk}{}_i{}^{l]} \,. \label{eq:4dQFlux}
\end{equation}
In particular, $\partial_i \Omega^{jkl}$ is not a spacetime tensor by itself but requires the $\hat{\Gamma}^{[jk}{}_i{}^{l]}$ term in order to transform covariantly. This flux transforms as
\begin{equation}
 Q_{i}{}^{jkl} \in \mbf{6}_2 \oplus \obf{10}_2 \,,
\end{equation}
of $\SL{4} \times \mathbb{R}^+$ and corresponds to those representations in \eqref{eq:SL5EmbeddingTensorSplit}.

\paragraph{$\tau$-flux:} The trace-part of $\hat{\Gamma}^{ij}{}_k{}^l$ is also a spacetime tensor as a result of the $\SL{5}$ section condition:
\begin{equation}
 \tau^{i,j} = \hat{\Gamma}^{ik}{}_k{}^j \,. \label{eq:4dTauFlux}
\end{equation}
This tensor can be split into a symmetric and an antisymmetric part and thus belongs to the representations
\begin{equation}
 \tau^{i,j} \in \mbf{6}_2 \oplus \mbf{10}_2 \,.
\end{equation}
Examples of compactifications with these fluxes turned on are given in \cite{Inverso:2017lrz}.

Comparing with \eqref{eq:SL5EmbeddingTensorSplit} we see that we have described all possible fluxes except for the $\mbf{1}_{-8}$, $\obf{20}_{-3}$ and $2 \cdot \obf{4}_{-3}$. These would correspond respectively to the four-form flux of M-theory $G_{ijkl} = 4 \partial_{[i} C_{jkl]} \in \mbf{1}_{-8}$, the trace and traceless parts of the geometric flux
\begin{equation}
 T_{ij}{}^k = 2 e_{\bi}{}^k \partial_{[i} e^{\bi}{}_{j]} \in \obf{4}_{-3} \oplus \obf{20}_{-3} \,,
\end{equation}
and a spacetime derivative of the determinant of the seven-dimensional metric, $\partial_i \Delta$, \cite{Blair:2014zba}.

\subsection{Five-dimensional locally non-geometric backgrounds} \label{s:SO55RFlux}
Our conventions for the $\SO{5,5}$ EFT, especially the generalised Lie derivative are summarised in appendix \ref{A:SO55}. We start with the $\SO{5,5}$ generalised vielbein in the non-geometric parameterisation, given by
\begin{equation}
 E_{\bar{M}}{}^{M}\equiv
 \begin{pmatrix}
 |e|^{-1 /4} e_{\bi}{}^i & 0  & 0  \\
 |e|^{-1 /4} \Omega^{\bi_1\bi_2 i} & |e|^{-1 /4} e^{\bi_1\bi_2}_{i_1i_2} & 0  \\
 -\frac1{12} |e|^{-1 /4} \epsilon_{k_1k_2k_3k_4k_5}\Omega^{ik_1 k_2}\Omega^{k_3k_4k_5} & -\frac1{6} |e|^{-1 /4} \epsilon_{i_1 i_2 k_1 k_2 k_3} \Omega^{k_1 k_2k_3} &|e|^{3 /4}   
 \end{pmatrix}\,,
\end{equation}
where
\begin{equation}
 e^{\bi\bj}_{ij} = e^{[\bi}{}_i e^{\bj]}{}_j \,.
\end{equation}
The factors of $|e|$ are dictated by the generalised Lie derivative, given that $e_{\bi}{}^i$ transforms as a spacetime vielbein, and similarly for the coefficients $-\frac1{12}$ and $-\frac16$. This differs from the parameterisation given in \cite{Lee:2016qwn} by the factors of $|e|$. However, these are necessary for the generalised vielbein to be a $\SO{5,5}$ group element (in the spinor representation).

From the generalised Lie derivative generated by a generalised vector of the form $V^M = \left( \xi^i,\, 0,\, 0 \right)$ given in \eqref{eq:SO55LieVielbein}, we find that the trivector transforms just as in the four-dimensional case \eqref{eq:SL5OmegaTransform} under spacetime diffeomorphisms:
\begin{equation}
 \begin{split}
  \delta_\xi \Omega^{ijk} &= \xi^l \partial_l \Omega^{ijk} - 3 \Omega^{l[ij} \partial_l \xi^{k]} - 3 \partial^{[ij} \xi^{k]} \,, \\
  &= L_\xi \Omega^{ijk} - 3 \partial^{[ij} \xi^{k]} \,, \label{eq:SO55OmegaTransform}
 \end{split}
\end{equation}
As before, we wish to construct a spacetime tensor from dual derivatives, now including $\partial^{ij}$ and $\partial^{ijklm} \equiv \epsilon^{ijklm} \partial^z$ of the trivector. To this end, we first introduce the improved dual derivatives
\begin{equation}
 \begin{split}
  \hpartial^{ij} &= |e|^{1/4} e_{\bi}{}^i e_{\bj}{}^j E^{\bi\bj\,M} \partial_M \\
  &= \partial^{ij} + \Omega^{ijk} \partial_k \,, \\
  \hpartial^z &= |e|^{1/4} E^{\bar{z}\,M} \partial_M \\
  &= |e| \partial^z - \frac16 \epsilon_{ijklm} \Omega^{ijk} \partial^{lm} - \frac1{12} \epsilon_{jklmn} \Omega^{ijk} \Omega^{lmn} \partial_i \,,
 \end{split}
\end{equation}
which satisfy
\begin{equation}
 \delta_\xi \hpartial^{ij} \phi = L_\xi \hpartial^{ij} \phi \,, \qquad \delta_\xi \hpartial^z \phi = L_\xi \hpartial^z \phi = \xi^i \partial_i \hpartial^z \phi \,,
\end{equation}
i.e. the improved dual derivatives of a scalar are spacetime tensors. This again makes use of the section condition
\begin{equation}
 \begin{split}
  \partial_i \otimes \partial^{ij} + \partial^{ij} \otimes \partial_i &= \partial_i \partial^{ij} = 0 \,, \\
  \partial_i \otimes \partial^z + \partial^z \otimes \partial_i + \frac12 \epsilon_{ijklm} \partial^{jk} \otimes \partial^{lm} &= \partial_i \partial^z + \frac14 \epsilon_{ijklm} \partial^{jk} \partial^{lm} = 0 \,,
 \end{split} \label{eq:SO55SectionCondition}
\end{equation}
with $\otimes$ denoting the action of the derivatives on two different fields.

Using these derivatives it is a straightforward, if lengthy, computation to construct a spacetime tensor from dual derivatives of the trivector $\Omega^{ijk}$. This leads to the unique combination
\begin{equation}
 R^{i,jklm} = 4 \hpartial^{i[j} \Omega^{klm]} - 2 e_{\bi}{}^{[j} \epsilon^{klm]in} \hpartial^z e^{\bi}{}_n \,. \label{eq:SO55RFlux}
\end{equation}
We see that the locally non-geometric flux in the five-dimensional compactification contains a new term compared to four dimensions \eqref{eq:SL5RFlux} involving the new dual derivative, associated with wrapping coordinates of M5-branes. Furthermore, unlike in four dimensions, $R^{i,jklm}$ contains a totally antisymmetric part, which we can view as a new locally non-geometric $R$-flux,
\begin{equation}
 R^{ijklm} \equiv 5 R^{[i,jklm]} = 20 \hpartial^{[ij} \Omega^{klm]} - 2 \epsilon^{ijklm} \hpartial^z |e| \,.
\end{equation}

Once again, we can also check this against the embedding tensor of six-dimensional maximal gauged SUGRAs. This takes values in the $\obf{16} \oplus \obf{144}$ of $\SO{5,5}$ which decomposes under $\SL{5} \times \mathbb{R}^+ \subset \SO{5,5}$ as
\begin{equation}
 \begin{split}
  \obf{16} &\longrightarrow \mbf{10}_1 \oplus \obf{5}_{-3} \oplus \mbf{1}_5 \,, \\
  \obf{144} &\longrightarrow \mbf{15}_1 \oplus \obf{45}_{-3} \oplus \mbf{24}_5 \oplus \mbf{5}_{-7} \oplus \mbf{40}_1 \oplus \mbf{10}_1 \oplus \obf{5}_{-3} \,.
 \end{split}
\end{equation}
From the branching of the $\mbf{16}$ we see also that the derivatives transform as
\begin{equation}
 \partial_i \in \obf{5}_{-3} \,, \qquad \partial^{ij} \in \mbf{10}_1 \,, \qquad \partial^{z} \in \mbf{1}_5 \,,
\end{equation}
while the adjoint decomposes as
\begin{equation}
 \mbf{45} \longrightarrow \mbf{24}_0 \oplus \mbf{10}_{-4} \oplus \obf{10}_4 \oplus \mbf{1}_0 \,,
\end{equation}
which allows us to identify that trivector as $\Omega^{ijk} \in \obf{10}_{4}$. 

Putting all this together we see that the $\mbf{24}_5 \in \mbf{144}$ corresponds to the mixed symmetry $R$-flux $R^{i,jklm} - R^{[i,jklm]}$ and the $\mbf{1}_5 \in \mbf{16}$ to the new totally antisymmetric $R$-flux $R^{ijklm}$. These are the only two fluxes with the right $\mathbb{R}^+$ charge to be locally non-geometric. However, while the mixed symmetry $R^{i,jklm}$ reduces to the NS-NS $R$-flux when considering an appropriate reduction on a circle, the new $R$-flux $R^{ijklm}$ does not. It will necessarily involve the R-R sector of string theory.

\subsubsection{Other non-geometric fluxes}
The other non-geometric fluxes are still exactly the same as in four dimensions, section \ref{s:SL5Other}. These are
\begin{equation}
 Q_i{}^{jkl} \in \mbf{10}_{1} \oplus \mbf{40}_{1} \,, \qquad \tau^{i,j} \in \mbf{10}_{1} \oplus \mbf{15}_{1} \,,
\end{equation}
with $Q_i{}^{jkl}$ as in \eqref{eq:4dQFlux} and $\tau^{i,j}$ as in \eqref{eq:4dTauFlux}. The remaining fluxes $\mbf{5}_{-7}$, $2 \cdot \obf{5}_{-3}$ and $\obf{45}_{-3}$ correspond to the four-form flux $G_{ijkl} = 4 \partial_{[i} C_{jkl]}$, the derivative of the determinant of the external six-dimensional metric, $\partial_i \Delta$, and the trace and traceless parts of the geometric flux $T_{ij}{}^k$, respectively.

\subsection{Six-dimensional locally non-geometric backgrounds} \label{s:E6RFlux}
We have summarised our conventions for the $\EG{6}$ EFT in appendix \ref{A:E6}. Following these conventions, the $\EG{6}$ generalised vielbein in the non-geometric parameterisation is given by
\begin{equation}
\small
E_{\bar{M}}{}^{M} = \begin{pmatrix}
|e|^{-1 /3} e_{\bi}{}^i & 0 & 0 \\
|e|^{-1 /3} \Omega^{\bi_1\bi_2 i} & |e|^{-1/3} e^{\bi_1\bi_2}_{i_1i_2} & 0  \\
\sqrt{\frac{5}{6}} |e|^{-1/3} \left(\Omega^{\bi_1 \bi_2 \bi_3 \bi_4 \bi_5 i} + \Omega^{[\bi_1 \bi_2 \bi_3 }\Omega^{\bi_4 \bi_5] i} \right) & \sqrt{\frac{10}{3}} |e|^{-1/3} e^{[\bi_1\bi_2}_{i_1 i_2} \Omega^{\bi_3 \bi_4 \bi_5]} &|e|^{-1/3} e^{\bi_1 \bi_2 \bi_3 \bi_4 \bi_5}_{i_1 i_2 i_3 i_4 i_5}
\end{pmatrix} \,, \normalsize
\end{equation}
where
\begin{equation}
 e^{\bi_1\bi_2\bi_3\bi_4\bi_5}_{i_1i_2i_3i_4i_5} = e^{[\bi_1}{}_{i_1} e^{\bi_2}{}_{i_2} e^{\bi_3}{}_{i_3} e^{\bi_4}{}_{i_4} e^{\bi_5]}{}_{i_5} \,,
\end{equation}
and $\Omega^{ijklmn}$ is the non-geometric dual of the usual 6-form flux of 11-dimensional supergravity. As previously, the factors of determinants and numerical coefficients come by requiring compatibility with the generalised Lie derivative and that $e_{\bi}{}^i$ transforms as a spacetime vielbein. If we act with the generalised Lie derivative using a generator of the form $V^M = \left( \xi^i,\, 0,\, 0 \right)$ corresponding to a spacetime diffeomorphism as in \eqref{eq:E6LieVielbein}, we find that $\Omega^{ijk}$ and $\Omega^{ijklmn}$ transform as
\begin{equation}
 \begin{split}
  \delta_\xi \Omega^{ijk} &= \xi^l \partial_l \Omega^{ijk} - 3 \Omega^{l[ij} \partial_l \xi^{k]} - 3 \partial^{[ij} \xi^{k]} \\
  &= L_\xi \Omega^{ijk} - 3 \partial^{[ij} \xi^{k]} \,, \\
  \delta_\xi \Omega^{ijklmn} &= \xi^p \partial_p \Omega^{ijklmn} - 6 \Omega^{p[ijklm} \partial_p \xi^{n]} + \frac{3}{5} \partial^{[ijklm} \xi^{n]} - 6 \Omega^{[ijk} \partial^{lm} \xi^{n]} \\
  &= L_\xi \Omega^{ijklmn} +\frac{3}{5} \partial^{[ijklm} \xi^{n]} - 6 \Omega^{[ijk} \partial^{lm} \xi^{n]} \,,
  \label{eq:E6OmegaTransform}
 \end{split}
\end{equation}
under spacetime diffeomorphisms.

Once again, we introduce improved dual derivatives
\begin{equation}
 \begin{split}
  \hpartial^{ij} &= |e|^{1/3} e_{\bi}{}^i e_{\bj}{}^i E^{\bi\bj\,M} \partial_M = \partial^{ij} + \Omega^{ijk} \partial_k \,, \\
  \hpartial^{ijklm} &= - \sqrt{5!}\, |e|^{1/3} e_{\bi}{}^i e_{\bj}{}^j e_{\bk}{}^k e_{\bl}{}^l e_{\bm}{}^m E^{\bi\bj\bk\bl\bm\,M} \partial_M \\
  &= \partial^{ijklm} - 20 \Omega^{[ijk} \partial^{lm]} - 10 \left( \Omega^{ijklmn} + \Omega^{[ijk} \Omega^{lm]n} \right) \partial_n \,,
 \end{split}
\end{equation}
where
\begin{equation}
 \partial_M = \left( \partial_i,\, \partial^{ij},\, - \frac{1}{\sqrt{5!}} \partial^{ijklm} \right) \,,
\end{equation}
so that $\hpartial^{ij} \phi$ and $\hpartial^{ijklm} \phi$ are spacetime tensors, where $\phi$ is a scalar. This works up to the section condition
\begin{equation}
 \begin{split}
  \partial_i \otimes \partial^{ij} + \partial^{ij} \otimes \partial_i &= \partial_i \partial^{ij} = 0 \,, \\
  \partial^{[ij} \otimes \partial^{kl]} + \frac{1}{12} \left( \partial_m \otimes \partial^{ijklm} + \partial^{ijklm} \otimes \partial_m \right) &= \partial^{[ij} \partial^{kl]} - \frac{1}{6} \partial_m \partial^{ijklm} = 0 \,, \\
  \partial^{i[j} \otimes \partial^{klmnp]} - \partial^{[klmnp} \otimes \partial^{j]i} &= \partial^{i[j} \partial^{klmnp]} = 0 \,,
 \end{split} \label{eq:E6SectionCondition}
\end{equation}
where $\otimes$ denotes the action of the derivatives on two different fields.

Using the improved dual derivatives, we find the same $R$-flux as in five dimensions
\begin{equation}
 \begin{split}
  R^{i,jklm} &= 4 \hpartial^{i[j} \Omega^{klm]} - 2 e_{\bi}{}^{[j} \hpartial^{klm]in} e^{\bi}{}_n \,,
 \end{split}
\end{equation}
which as in five dimensions can be decomposed into a totally antisymmetric part $R^{ijklm}$ and a mixed symmetry one, as well as a new $R$-flux involving $\Omega^{ijklmn}$, given by
\begin{equation}
 R^{ij} = \frac1{5!} \epsilon_{klmnpq} \hpartial^{klmnp} \Omega^{qij} + \frac1{72} \epsilon_{klmnpq} \hpartial^{ij} \Omega^{klmnpq} + \frac1{36} \epsilon_{klmnpq} \Omega^{klm} \hpartial^{ij} \Omega^{npq} \,. \label{eq:E6RFlux}
\end{equation}
Again, we can check that we have found all of the $R$-fluxes by comparison with the $\EG{6}$ embedding tensor, which transforms under the $\obf{27} \oplus \obf{351}$ of $\EG{6}$. We decompose $\EG{6} \longrightarrow \SL{6} \times \mathbb{R}^+$ and find that the embedding tensor branches as
\begin{equation}
 \begin{split} 
  \obf{27}& \rightarrow \mbf{15}_{0} \oplus \obf{6}_2 \oplus \obf{6}_{-2} \,,\\
  \obf{351} &\rightarrow \mbf{21}_{0} \oplus \obf{84}_{2} \oplus \obf{84}_{-2} \oplus \mbf{105}_{0} \oplus \mbf{15}_4 \oplus \mbf{15}_0 \oplus \mbf{15}_{-4} \oplus \obf{6}_2 \oplus \obf{6}_{-2} \,, \label{eq:E6EmbeddingTensorBranch}
 \end{split}
\end{equation}
and the adjoint as
\begin{equation}
 \mbf{78} \longrightarrow \mbf{35}_0 \oplus \mbf{20}_2 \oplus \mbf{20}_{-2} \oplus \mbf{1}_4 \oplus \mbf{1}_{0} \oplus \mbf{1}_{-4} \,. \label{eq:E6AdjointBranch}
\end{equation}
From this we identify $\Omega^{ijk} \in \mbf{20}_2$ and $\Omega^{ijklmn} \in \mbf{1}_4$, while the branching of the $\mbf{27}$ in \eqref{eq:E6EmbeddingTensorBranch} shows that the derivatives transform as
\begin{equation}
 \partial_i \in \obf{6}_{-2} \,, \qquad \partial^{ij} \in \mbf{15}_{0} \,, \qquad \partial^{ijklm} \in \obf{6}_2 \,.
\end{equation}
We can now construct locally non-geometric fluxes by acting with either of the dual derivatives on $\Omega^{ijk}$ or $\Omega^{ijklmn}$. Thus, the locally non-geometric fluxes must have $\mathbb{R}^+$ charge $+2$, $+4$ or $+6$. From the embedding tensor we see that there are therefore only three locally non-geometric fluxes $\obf{6}_2 \subset \mbf{27}$ and $\obf{84}_2 \subset \mbf{351}$ and $\mbf{15}_4 \subset \mbf{351}$ which correspond exactly to
\begin{equation}
 R^{i,jklm} \in \obf{6}_2 \oplus \obf{84}_2 \,, \qquad R^{ij} \in \mbf{15}_4 \,.
\end{equation}
The final $\obf{6}_2 \subset \mbf{351}$ in the embedding tensor corresponds to a globally non-geometric $Q$-flux \cite{Lee:2016qwn}:
\begin{equation}
 Q_i{}^{jklmnp} = \partial_i \Omega^{jklmnp} + 2 \Omega^{[jkl} \partial_i \Omega^{mnp]} - \frac35 \Gamma^{[jklmn}{}_i{}^{p]} \,,
\end{equation}
where we define $\Gamma^{[jklmn}{}_i{}^{p]}$ below in subsection \ref{s:E6OtherFluxes}. This confirms that we have constructed all possible locally non-geometric fluxes in six dimensions.

\subsubsection{Other non-geometric fluxes} \label{s:E6OtherFluxes}
There are other spacetime tensors that can be constructed in the non-geometric parameterisation, i.e. in terms of the fields $g_{ij}$, $\Omega^{ijk}$ and $\Omega^{ijklmn}$. These include the geometric flux
\begin{equation}
 T_{ij}{}^k = 2 e_{\bi}{}^k \partial_{[i} e^{\bi}{}_{j]} \,,
\end{equation}
the derivative of the determinant of the external metric $\partial_i \Delta$, as well as other non-geometric fluxes involving either spacetime derivatives of $\Omega^{ijk}$ and $\Omega^{ijklmn}$, or dual derivatives of the vielbein.

We begin by introducing the connection-like objects
\begin{equation}
 \begin{split}
  \Gamma_{ij}{}^{k} &= e_{\bi}{}^{k} \partial_i e^{\bi}{}_{j} \,, \\
  \hat{\Gamma}^{ij}{}_k{}^l &= e_{\bi}{}^{l} \hpartial^{ij} e^{\bi}{}_{k} \,, \\
  \hat{\Gamma}^{ijklm}{}_n{}^p &= e_{\bi}{}^p \hpartial^{ijklm} e^{\bi}{}_n \,.  
 \end{split}
\end{equation}
$\Gamma_{ij}{}^k$ is the Weitzenb\"ock connection which is natural on parallelisable backgrounds. These will appear inside the embedding tensor in combinations that are well-defined on generalised parallelisable spaces. Under spacetime diffeomorphisms the above objects transform as
\begin{equation}
 \begin{split}
  \delta_\xi \Gamma_{ij}{}^k &= L_\xi \Gamma_{ij}{}^k + \partial_i \partial_j \xi^k \,, \\
  \delta_\xi \hat{\Gamma}^{ij}{}_k{}^l &= L_\xi \hat{\Gamma}^{ij}{}_k{}^l + \hpartial^{ij} \partial_k \xi^l \,, \\
  \delta_\xi \hat{\Gamma}^{ijklm}{}_n{}^p &= L_\xi \hat{\Gamma}^{ijklm}{}_n{}^p + \hpartial^{ijklm} \partial_n \xi^p \,.
 \end{split}
\end{equation}

\paragraph{$Q$-fluxes:} Using these, we once again find the $Q$-flux associated to $\Omega^{ijk}$, as in four and five dimensions,
\begin{equation}
 Q_i{}^{jkl} = \partial_i \Omega^{jkl} + 3 \hat{\Gamma}^{[jk}{}_i{}^{l]} \,,
\end{equation}
as well as a new $Q$-flux associated to $\Omega^{ijklmn}$,
\begin{equation}
 Q_i{}^{jklmnp} = \partial_i \Omega^{jklmnp} + 2 \Omega^{[jkl} \partial_i \Omega^{mnp]} - \frac35 \hat{\Gamma}^{[jklmn}{}_i{}^{p]} \,.
\end{equation}
The first two terms of this $Q$-flux were also constructed in \cite{Lee:2016qwn}, where the combination was called $S_i{}^{jklmnp}$. However \cite{Lee:2016qwn} assumed the simplifying assumptions that $\hat{\partial}^{ij} = \hat{\partial}^{ijklm} = 0$ so that the $\hat{\Gamma}^{ij}{}_k{}^l$ and $\hat{\Gamma}^{ijklm}{}_n{}^p$ terms of the $Q$-flux were not captured.

\paragraph{$\tau$-flux:} The only other non-geometric flux that we can construct is
\begin{equation}
 \tau^{i,j} = \hat{\Gamma}^{ik}{}_k{}^j \,,
\end{equation}
which is unchanged from the four-dimensional case \eqref{eq:4dTauFlux}.

These tensors transform in the following representations of $\SL{6} \times \mathbb{R}^+$
\begin{equation}
 Q_{i}{}^{jkl} \in \mbf{15}_0 \oplus \mbf{105}_0 \,, \qquad Q_i{}^{jklmnp} \in \obf{6}_2 \,, \qquad \tau^{i,j} \in \mbf{15}_0 \oplus \mbf{21}_0 \,,
\end{equation}
while the geometric flux and derivative of the determinant of the external metric transform as
\begin{equation}
 T_{ij}{}^k \in \obf{6}_{-2} \oplus \obf{84}_{-2} \,, \qquad \partial_i \Delta \in \obf{6}_{-2} \,.
\end{equation}
The only remaining unaccounted flux in \eqref{eq:E6EmbeddingTensorBranch} is given by the four-form field strength $G_{ijkl} = 4 \partial_{[i} C_{jkl]} \in \mbf{15}_{-4}$ which vanishes in the non-geometric parameterisation. This shows that we have identified all the non-geometric fluxes.

\subsection{Seven-dimensional locally non-geometric backgrounds} \label{s:E7RFlux}
Appendix \ref{A:E7} summarises our conventions for the $\EG{7}$ EFT. To write down the $\EG{7}$ generalised vielbein in non-geometric parameterisation, it is worthwhile to introduce
\begin{equation}
 \begin{split}
  V_{ijkl} &= \epsilon_{ijklmnp} \Omega^{mnp} \,, \\
  W_{ij}{}^k &= V_{ijlm} \Omega^{klm} \,, \\
  X_{ij}{}^k &= \epsilon_{ijlmnpq} \Omega^{lmnpqk} \,.
 \end{split}
\end{equation}
In terms of these objects, the non-geometrically parameterised $\EG{7}$ generalised vielbein is given by
\begin{small}
\begin{equation}
 E_{\bar{M}}{}^M = \begin{pmatrix}
  |e|^{-1/2}e_{\bi}{}^i & 0  & 0 & 0 \\
  |e|^{-1/2} \Omega^{\bi_1\bi_2 i} & |e|^{-1/2} e^{\bi_1\bi_2}_{i_1 i_2} & 0 & 0 \\
  \frac{1}{24} |e|^{-1/2} \left( X_{\bi_1\bi_2}{}^i + W_{\bi_1\bi_2}{}^i \right) & \frac1{12} |e|^{-1/2} V_{\bi_1\bi_2 i_1 i_2} & |e|^{1/2} e_{\bi_1\bi_2}^{i_1 i_2} & 0 \\
 |e|^{-1/2} \frac1{144} \left( W_{jk}{}^i \Omega^{\bi jk} - 3 X_{jk}{}^i \Omega^{\bi jk} \right) & 
  \frac1{48} |e|^{-1/2} \left( W_{i_1i_2}{}^{\bi} - X_{i_1i_2}{}^{\bi} \right) &
  \frac{1}{2}|e|^{1/2} \Omega^{\bi i_1 i_2} & |e|^{1/2} e^{\bi}{}_i \end{pmatrix}\,. \label{eq:E7GenVielbein}
\end{equation}
\end{small}

From the action of the generalised Lie derivative on the above generalised vielbein \eqref{eq:E7GenVielbein}, with a generalised vector field $V^M = \left( \xi^i,\, 0,\, 0,\, 0 \right)$ corresponding to a spacetime diffeomorphism as in \eqref{eq:E7LieVielbein}, we find the following transformation laws
\begin{equation}
 \begin{split}
  \delta_\xi \Omega^{ijk} &= L_\xi \Omega^{ijk} - 3 \partial^{[ij} \xi^{k]} \,, \\
  \delta_\xi \Omega^{ijklmn} &= L_\xi \Omega^{ijklmn} - \frac1{10} \epsilon^{ijklmnp} \partial_{pq} \xi^q - 6 \Omega^{[ijk} \partial^{lm} \xi^{n]} \,, 
 \end{split}
\end{equation}
under spacetime diffeomorphisms. Here we have defined
\begin{equation}
 \begin{split}
  L_\xi \Omega^{ijk} &= \xi^l \partial_l \Omega^{ijk} - 3 \Omega^{l[ij} \partial_l \xi^{k]} \,, \\
  L_\xi \Omega^{ijklmn} &= \xi^p \partial_p \Omega^{ijklmn} - 6 \Omega^{p[ijklm} \partial_p \xi^{n]} \,.
 \end{split}
\end{equation}

Before constructing the locally non-geometric fluxes, we first introduce the improved dual derivatives
\begin{equation}
 \begin{split}
  \hpartial^{ij} &= |e|^{1/2} e_{\bi}{}^i e_{\bj}{}^i E^{\bi\bj\,M} \partial_M = \partial^{ij} + \Omega^{ijk} \partial_k \,, \\
  \hpartial_{ij} &= -2 \, |e|^{1/2} e^{\bi}{}_i e^{\bj}{}_j E_{\bi\bj}{}^{M} \partial_M \\
  &= |e| \partial_{ij} - \frac16 \epsilon_{ijklmnp} \Omega^{klm} \partial^{np} - \frac1{12} \epsilon_{ijklmnp} \left( \Omega^{klmnpq} + \Omega^{klm} \Omega^{npq} \right) \partial_q \,, \\
  \hpartial^i &= |e|^{1/2} e_{\bi}{}^i E^{\bi\,M} \partial_M \\
  &= |e| \partial^i - \frac14 |e| \Omega^{ijk} \partial_{jk} + \frac1{48} \epsilon_{jklmnpq} \left( \Omega^{jkl} \Omega^{imn} - \Omega^{jklmni} \right) \partial^{pq} \\
  & \quad + \frac1{144} \epsilon_{jklmnpq} \Omega^{pqr} \left( \Omega^{jkl} \Omega^{imn} - 3 \Omega^{jklmni} \right) \partial_r \,, \label{eq:E7HatDerivatives}
 \end{split}
\end{equation}
where
\begin{equation}
 \partial_M = \left( \partial_i,\, \partial^{ij} ,\, -\frac12 \partial_{ij} ,\, \partial^i \right) \,,
\end{equation}
and $\partial_{ij} \equiv \frac1{5!} \epsilon_{ijklmnp} \partial^{klmnp}$, while $\partial^i \equiv \frac1{7!} \epsilon_{jklmnpq} \partial^{jklmnpq,i}$. The improved dual derivative of a scalar field $\phi$ is a spacetime tensor, i.e.
\begin{equation}
 \delta_\xi \hpartial^{ij} \phi = L_\xi \hpartial^{ij} \phi \,, \qquad \delta_\xi \hpartial_{ij} \phi = L_\xi \hpartial_{ij} \phi \,, \qquad \delta_\xi \hpartial^i \phi = L_\xi \hpartial^i \phi \,,
\end{equation}
up to the section condition
\begin{equation}
 \begin{split}
  \partial_i \otimes \partial^{ij} + \partial^{ij} \otimes \partial_i = \partial_i \partial^{ij} &= 0 \,, \\
  \partial^i \otimes \partial_{ij} + \partial_{ij} \otimes \partial^{ij} = \partial^i \partial_{ij} &= 0 \,, \\
  \partial_i \otimes \partial^i = \partial_i \partial^i &= 0 \,, \\
  \partial^{ij} \otimes \partial_{ij} = \partial_{ij} \partial^{ij} &= 0 \,, \\
  \partial_i \otimes \partial^j + \partial^j \otimes \partial_i - \frac12 \partial_{ik} \otimes \partial^{jk} - \frac12 \partial^{jk} \otimes \partial_{ik} = \partial_i \partial^j - \frac12 \partial_{ik} \partial^{jk} &= 0 \,, \\
  \partial_{[ij} \otimes \partial_{kl]} - \frac13 \epsilon_{ijklmnp} \left( \partial^{mn} \otimes \partial^p + \partial^p \otimes \partial^{mn} \right) = \partial_{[ij} \partial_{kl]} - \frac23 \epsilon_{ijklmnp} \partial^{mn} \partial^p &= 0 \,, \\
  \partial_{[ij} \otimes \partial_{k]} + \partial_{[i} \otimes \partial_{jk]} + \frac16 \epsilon_{ijklmnp} \partial^{mn} \otimes \partial^{np} = \partial_{[ij} \partial_{k]} + \frac{1}{12} \epsilon_{ijklmnp} \partial^{mn} \partial^{np} &= 0 \,, \label{eq:E7SectionCondition}
 \end{split}
\end{equation}
where again $\otimes$ denotes the derivatives acting on two different fields.

Using the improved dual derivatives we find the following locally non-geometric fluxes.
\begin{equation}
 \begin{split}
  R^{i,jklm} &= 4 \hpartial^{i[j} \Omega^{klm]} - e_{\bi}{}^{[j} \epsilon^{klm]inpq} \hpartial_{pq} e^{\bi}{}_n \,, \\
  R^{ij}{}_k &= \hpartial_{kl} \Omega^{ijl} - \frac1{72} \epsilon_{klmnpqr} \hpartial^{ij} \Omega^{lmnpqr} - \frac{1}{36} \epsilon_{klmnpqr} \Omega^{lmn} \hpartial^{ij} \Omega^{pqr} \\
  & \quad + 4 e_{\bi}{}^{[i} \hpartial^{j]} e^{\bi}{}_k + 4 \delta^{[i}_k e_{\bi}{}^{j]} \hpartial^l e^{\bi}{}_l \,, \\
  R^i &= \hpartial_{jk} \Omega^{ijk} - 4 \hpartial^i |e| - 8 e_{\bi}{}^i \hpartial^j e^{\bi}{}_j \,, \\
  R &= \epsilon_{ijklmnp} \hpartial^{i} \Omega^{jklmnp} - 2 \epsilon_{ijklmnp} \Omega^{ijk} \hpartial^m \Omega^{lnp} \,, \\
  R^{ijkl} &= \frac58 \hpartial^{[i} \Omega^{jkl]} + \frac12 \hpartial_{pq} \Omega^{pqijkl} + \frac14 \Omega^{[ijk} \hpartial_{pq} \Omega^{l]pq} \,,
 \end{split}
\end{equation}
where the first line corresponds to the $R$-flux as already seen in five-dimensional compactifications, the second and third line represent modifications of the $R$-flux that first arose in six-dimensional compactifications and the final two locally non-geometric fluxes $R$ and $R^{ijkl}$ are new and only arise in compactifications of seven or more dimensions. In fact, because seven-dimensional compactifications can depend on a Kaluza-Klein monopole winding number $y_i$, there is a dual derivative $\partial^i$ in the $\mbf{7}$ of $\SL{7}$. This means that the new locally non-geometric fluxes mirror the usual $G_4$ and $G_7$ fluxes of supergravity.

We can also show that these are the only locally non-geometric fluxes by comparison with the embedding tensor of four-dimensional maximal gauged SUGRAs, which takes values in the $\mbf{56} \oplus \mbf{912}$ of $\EG{7}$. Decomposing $\EG{7} \longrightarrow \SL{7} \times \mathbb{R}^+$, we find
\begin{equation}
 \begin{split}
  \mbf{56} &\longrightarrow \obf{7}_{-6} \oplus \mbf{21}_{-2} \oplus \obf{21}_{2} \oplus \mbf{7}_6 \,, \\
  \mbf{912} &\longrightarrow \mbf{1}_{-14} \oplus \mbf{35}_{-10} \oplus \obf{140}_{-6} \oplus \obf{7}_{-6} \oplus \mbf{224}_{-2} \oplus \mbf{21}_{-2} \oplus \mbf{28}_{-2} \\
  & \quad \oplus \obf{28}_{2} \oplus \obf{21}_{2} \oplus \obf{224}_{2} \oplus \mbf{7}_{6} \oplus \mbf{140}_{6} \oplus \obf{35}_{10} \oplus \mbf{1}_{14} \,.
 \end{split} \label{eq:E7EmbeddingTensorSplit}
\end{equation}
From the decomposition of the $\mbf{56}$ we also see that the derivatives transform as
\begin{equation}
 \partial_i \in \obf{7}_{-6} \,, \qquad \partial^{ij} \in \mbf{21}_{-2} \,, \qquad \partial_{ij} \in \obf{21}_{2} \,, \qquad \partial^i \in \mbf{7}_6 \,.
\end{equation}
The decomposition of the adjoint representation of $\EG{7}$ shows that
\begin{equation}
 \Omega^{ijk} \in \mbf{35}_4 \,, \qquad \Omega^{ijklmn} \in \obf{7}_8 \,.
\end{equation}
We see that the locally non-geometric fluxes correspond to
\begin{equation}
 \begin{split}
  R^{i,jklm} &\in \obf{21}_2 \oplus \obf{224}_2 \,, \qquad R^{ij}{}_k \in \mbf{7}_6 \oplus \mbf{140}_6 \,, \qquad R^i \in \mbf{7}_6 \,, \\
  & \qquad \qquad \quad R^{ijkl} \in \obf{35}_{10} \,, \qquad R \in \mbf{1}_{14} \,.
 \end{split}
\end{equation}

\subsubsection{Other non-geometric fluxes} \label{s:E7OtherFluxes}
The other non-geometric fluxes are exactly as in the six-dimensional case, explained in section \ref{s:E6OtherFluxes}. These transform in the representations
\begin{equation}
 Q_{i}{}^{jkl} \in \mbf{21}_{-2} \oplus \mbf{224}_{-2} \,, \qquad Q_i{}^{jklmnp} \in \obf{21}_2 \oplus \obf{28}_2 \,, \qquad \tau^{i,j} \in \mbf{21}_{-2} \oplus \mbf{28}_{-2} \,,
\end{equation}
and the geometric flux transforms in
\begin{equation}
 T_{ij}{}^k \in \obf{7}_{-6} \oplus \obf{140}_{-6} \,.
\end{equation}
We also have $\partial_i \Delta \in \obf{7}_{-6}$, the derivative of the determinant of the external metric. The remaining fluxes in \eqref{eq:E7EmbeddingTensorSplit} correspond to the four-form and seven-form flux $G_{ijkl} = 4 \partial_{[i} C_{jkl]} \in \mbf{35}_{-10}$ and $G_{ijklmnp} = 7 \partial_{[i} C_{jklmnp]} - 35 C_{[ijk} G_{lmnp]} \in \mbf{1}_{-14}$. Thus, we have found all the non-geometric fluxes.

\section{Duality chains and missing momenta} \label{s:DualityChain}

\subsection{Geometric flux}
We will now construct examples of new locally non-geometric backgrounds by acting with U-dualities on parallelisable spaces with ``geometric flux''. For simplicity, we will refer to these geometric spaces as ``twisted tori'' even though some examples are non-compact. In \cite{Gunaydin:2016axc}, it was shown that a particular example, the direct product of the Heisenberg Nilmanifold with a circle, does not allow for certain membrane wrapping states. After applying U-dualities, some of these states would have become momentum modes in the locally non-geometric background, which as a result must be missing. To understand how this generalises to locally non-geometric background with the new $R$-fluxes considered here, we begin with a detailed look at two kinds of spaces with geometric flux that will be the starting point of our duality chains.

\subsubsection{Nilmanifold} \label{s:TwistedTorus}
The first is the Heisenberg Nilmanifold, ${\cal N}_3$, which be defined as the coset space \cite{Kachru:2002sk}
\begin{equation}
 {\cal N}_3 = \frac{G_N(\mathbb{R})}{G_N(\mathbb{Z})} \,,
\end{equation}
where $G_N(\mathbb{R})$ and $G_N(\mathbb{Z})$ are the matrix groups
\begin{equation}
 G_N(\mathbb{F}) = \left\{ \left. \begin{pmatrix}
  1 & c & - \frac{1}{N}\,a \\
  0 & 1 & b \\
  0 & 0 & 1
 \end{pmatrix} \,\, \right\vert \quad a,\, b,\, c \in \mathbb{F}\,\, \right\} \,,
\end{equation}
and $\mathbb{F} = \mathbb{R}$ or $\mathbb{F} = \mathbb{Z}$, respectively.\footnote{The terminology ``Nilmanifold'' refers to the fact that the Lie algebra of $G_N(\mathbb{R})$ is nilpotent.} For fixed $N$, $G_N(\mathbb{R})$ is isomorphic to the continuous Heisenberg group, which is topologically $\mathbb{R}^3$. $G_N(\mathbb{Z})$ are different discrete subgroups of $G_N(\mathbb{R})$, with $G_1(\mathbb{Z})$ corresponding to the discrete Heisenberg group. If we introduce coordinates $\x$, $\y$ and $\z$ on this coset space by taking
\begin{equation}
 G_N(\mathbb{R}) = \left\{ \left. \begin{pmatrix}
  1 & \z & - \frac{1}{N}\,\x \\
  0 & 1 & \y \\
  0 & 0 & 1
 \end{pmatrix} \,\, \right\vert \quad \x,\, \y,\, \z \in \mathbb{R}\,\, \right\} \,,
\end{equation}
then the quotient leads to the identifications
\begin{equation}
 \left(\x,\,\y,\,\z\right) \sim \left(\x+1,\,\y,\,\z\right) \sim \left(\x,\,\y,\,\z+1\right) \sim \left(\x-N\z,\,\y+1,\,\z\right) \,.
\end{equation}

Key for us is that on ${\cal N}_3$ we can introduce three well-defined 1-forms given by
\begin{equation}
 \begin{split}
  e^{\bar{1}} &= \dx + N \y \dz \,, \\
  e^{\bar{2}} &= \dy \,, \\
  e^{\bar{3}} &= \dz \,.
 \end{split} \label{eq:TT1forms}
\end{equation}
Clearly, ${\cal N}_3$ is a parallelisable space. The 1-forms satisfy
\begin{equation}
 de^{\bar{1}} = N e^{\bar{2}} \wedge e^{\bar{3}} \,, \qquad de^{\bar{2}} = de^{\bar{3}} = 0 \,,
\end{equation}
so that the ``geometric flux'', which is defined as
\begin{equation}
 de^{\bi} = T_{\bj\bk}{}^{\bi} e^{\bj} \wedge e^{\bk} \,, \label{eq:GeometricFlux}
\end{equation}
is given by
\begin{equation}
 T_{23}{}^1 = N \,.
\end{equation}

In the upcoming duality chains, we will be considering spaces that are direct product of tori with ${\cal N}_3$, which we label as
\begin{equation}
 {\cal N}_3^m \equiv {\cal N}_3 \times T^m \,.
\end{equation}
We will use $\left( x^4,\, \ldots,\, x^{m+3}\right)$ as the usual coordinates on the $T^m$. We will also consider a metric on ${\cal N}_3^m$ constructed from the well-defined 1-forms on ${\cal N}_3$ given in equation \eqref{eq:TT1forms},
\begin{equation}
 \begin{split}
  ds^2 &= \left( \dx + N \y \dz \right)^2 + \pdy^2 + \pdz^2 + \pdw^2 + \ldots + \left( dx^{m+3} \right)^2 \,.
 \end{split}
\end{equation}

\subsubsection{Non-unimodular geometric flux}
The second space with geometric flux that we will consider is a two-dimensional manifold, ${\cal N}_2$, with globally well-defined 1-forms
\begin{equation}
 e^{\bar{1}} = d\x + N \x \dy \,, \qquad e^{\bar{2}} = \dy \,. \label{eq:Solv1Forms}
\end{equation}
These satisfy
\begin{equation}
 de^{\bar{1}} = N e^{\bar{1}} \wedge e^{\bar{2}} \,, \qquad de^{\bar{2}} = 0 \,,
\end{equation}
so that ${\cal N}_2$ has non-unimodular geometric flux \eqref{eq:GeometricFlux}
\begin{equation}
 T_{12}{}^1 = N \,.
\end{equation}
Note that the second deRham cohomology class of this space vanishes
\begin{equation}
 H^2_{dR}\left({\cal N}_2\right) = 0 \,. \label{eq:2ndCohoClassN2}
\end{equation}
This implies that ${\cal N}_2$ cannot be compact. If it were, then its second homology class would necessarily be $H_2\left({\cal N}_2,\mathbb{Z}\right) = \mathbb{Z}$, since ${\cal N}_2$ is clearly orientable. However, this would also imply $H^2\left({\cal N}_2,\mathbb{R}\right) = \mathbb{R}$ which is in contradiction with \eqref{eq:2ndCohoClassN2}.

An example of such a space is given by the solvmanifold
\begin{equation}
 {\cal S}_2 = \frac{S_N(\mathbb{R})}{\Lambda} \,,
\end{equation}
where $S_N(\mathbb{R})$ is the matrix group
\begin{equation}
 S_N(\mathbb{R}) = \left\{ \left. \begin{pmatrix}
  e^{\y} & e^{(1+N)\y}\, \x \\ 0 & e^{(1+N)\y} 
 \end{pmatrix} \right\vert \left(\x,\, \y\right) \in \mathbb{R}^2 \right\} \,.
\end{equation}
The discrete subgroup $\Lambda$ is defined as
\begin{equation}
 \Lambda = \left\{ \left. \begin{pmatrix}
   1 & b \\ 0 & 1 
  \end{pmatrix} \right\vert b \in \mathbb{Z} \right\} \,,
\end{equation}
so that the right-quotient of $S_N(\mathbb{R})$ by $\Lambda$ leads to the identifications
\begin{equation}
 \left(\x, \y \right) \sim \left( \x + e^{-N\y} , \y \right) \,.
\end{equation}
We see that ${\cal S}_2$ is an infinitely-long cylinder $S^1 \times \mathbb{R}$ where the radius of the cylinder increases along the length of the cylinder. It is worthwhile to mention that because ${\cal S}_2$ is non-compact, we cannot differentiate topologically between non-zero values of $N$. The distinction between them only arises once we introduce a length scale on $\x$, e.g. through a metric.

In the following, we will consider the spaces ${\cal N}_2^m \equiv {\cal N}_2 \times T^m$, and we will use $\left(\z, \ldots, x^{m+2}\right)$ as the usual coordinates on $T^m$. The metric on these spaces will be given by
\begin{equation}
 ds^2 = \left( \dx + N \x \dy \right)^2 + \pdy^2 + \pdz^2 + \ldots + \left(dx^{m+2}\right)^2 \,.
\end{equation}

\subsection{Wrapping states}
We need to compute the second and fifth homology groups of these spaces and, in the seven-dimensional case, identify the $S^1$'s in order to understand what possible M2, M5, and KK monopole wrapping modes are allowed. This will tell us what momenta are missing in the dual locally non-geometric background.

\subsubsection{Wrapping states of ${\cal N}_3^1$} \label{s:N31}
For ${\cal N}_3^1$ we use the fact that
\begin{equation}
 e^{\bar{2}} \wedge e^{\bar{3}} = \frac1{N} de^{\bar{1}} \,, \qquad d\left( e^{\bar{1}} \wedge e^{\bar{4}} \right) = N e^{\bar{2}} \wedge e^{\bar{3}} \wedge e^{\bar{4}} \,,
\end{equation}
to deduce that the second de Rham cohomology is
\begin{equation}
 H^2_{dR}({\cal N}_3^1) = \mathbb{R} \oplus \mathbb{R} \oplus \mathbb{R} \oplus \mathbb{R} \,. \label{eq:N31deRham}
\end{equation}
However, to identify possible membrane wrapping modes, we need to compute the integer homology groups. From \eqref{eq:N31deRham} we see that
\begin{equation}
 H_2({\cal N}_3^1,\mathbb{Z}) = \mathbb{Z} \oplus \mathbb{Z} \oplus \mathbb{Z} \oplus \mathbb{Z} \oplus \mathbb{Z}_{k_1} \oplus \mathbb{Z}_{k_2} \,,
\end{equation}
with $k_1$ and $k_2$ integers determining the rank of the torsion subgroups generated by the $\left(\y\z\right)$ and $\left(\x\w\right)$ cycles. However, because the ${\cal N}_3$ submanifold is closed and oriented, $H_2\left({\cal N}_3\right)$ cannot have torsion and therefore M2-branes cannot wrap the two-cycle $\left(\y\z\right)$. Hence $w^{23} = 0$. As discussed in \cite{Gunaydin:2016axc}, this missing wrapping mode can also be understood from the Freed-Witten anomaly in the IIB background obtained by dualising along $\x$ and $\w$. Thus ${\cal N}_3^1$ provides a geometric realisation of the Freed-Witten anomaly of the $T^3$ with H-flux.

\subsubsection{Wrapping states of ${\cal N}_3^3$} \label{s:N33}
In section \ref{s:E66Chain} we will consider a duality chain based on the six-dimensional manifold ${\cal N}_3^3$. Thus, we need to understand its possible membrane and five-brane wrapping modes. We begin by computing the de Rham cohomology. Using that
\begin{equation}
 \begin{split}
  de^{\bar{1}} = N e^{\bar{2}} \wedge e^{\bar{3}} \,,
 \end{split}
\end{equation}
we find that the following are not 2- and 4-forms are not closed
\begin{equation}
 \begin{split}
  d\left(e^{\bar{1}} \wedge e^{\bar{4}} \right) &= d\left(e^{\bar{1}} \wedge e^{\bar{5}} \right) = d\left(e^{\bar{1}} \wedge e^{\bar{6}} \right) \neq 0 \,, \\
  d\left( e^{\bar{1}} \wedge e^{\bar{4}} \wedge e^{\bar{5}} \wedge e^{\bar{6}} \right) &= N e^{\bar{2}} \wedge e^{\bar{3}} \wedge e^{\bar{4}} \wedge e^{\bar{5}} \wedge e^{\bar{6}} \neq 0 \,.
 \end{split}
\end{equation}
The last equation implies that the five-form $e^{\bar{2}} \wedge e^{\bar{3}} \wedge e^{\bar{4}} \wedge e^{\bar{5}} \wedge e^{\bar{6}}$ is exact and therefore trivial in de Rham cohomology.

The above relations in particular imply that the two-cycles $\left(\y\z\right)$ and $\left(\y\z\w\xf\xsi\right)$ are either homologically trivial or generate torsion subgroups of $H_2\left({\cal N}_3\right) \subset H_2\left({\cal N}_3^3\right)$ and $H_5\left({\cal N}_3^3\right)$, respectively. However, because both ${\cal N}_3$ and ${\cal N}_3^3$ are closed and oriented, $H_2\left({\cal N}_3,\mathbb{Z}\right)$ and $H_5\left({\cal N}_3^3,\mathbb{Z}\right)$ cannot have torsion subgroups. Therefore there are no membrane or five-brane wrapping modes associated with the $\left(\y\z\right)$ and $\left(\y\z\w\xf\xsi\right)$ cycles, and $w^{23} = w^{23456} = 0$. Just as in ${\cal N}_3^1$, these missing wrapping modes can equivalently be understood as being due to the Freed-Witten anomaly of the IIB background obtained by dualising along $\x$ and $\xsi$.

\subsubsection{Wrapping states of ${\cal N}_3^4$} \label{s:N34}
We now consider the seven-dimensional manifold ${\cal N}_3^4 = {\cal N}_3 \times T^4$, which we will dualise in \ref{s:E7Chain} to obtain the new locally non-geometric fluxes in seven dimensions. Using the notation of \eqref{eq:TT1forms} and letting
\begin{equation}
 e^{\bfo} = \dw \,, \qquad e^{\bfi} = d\xf \,, \qquad e^{\bar{6}} = d\xsi \,, \qquad e^{\bar{7}} = d\xse \,.
\end{equation}
we find
\begin{equation}
 \begin{split}
  e^{\bar{2}} \wedge e^{\bar{3}} &= \frac1N de^{\bar{1}} \,, \\
  d\left(e^{\bar{1}} \wedge e^{\bar{\mu}} \right) & \neq 0 \,, \\
  e^{\bar{2}} \wedge e^{\bar{3}} \wedge e^{\bar{\mu}} \wedge e^{\bar{\nu}} \wedge e^{\bar{\rho}} &= \frac1N d\left( e^{\bar{1}} \wedge e^{\bar{\mu}} \wedge e^{\bar{\nu}} \wedge e^{\bar{\rho}} \right) \,, \\
  d \left( e^{\bar{1}} \wedge e^{\bar{4}} \wedge e^{\bar{5}} \wedge e^{\bar{6}} \wedge e^{\bar{7}} \right) &= N e^{\bar{2}} \wedge e^{\bar{3}} \wedge e^{\bar{4}} \wedge e^{\bar{5}} \wedge e^{\bar{6}} \wedge e^{\bar{7}} \,,
 \end{split}
\end{equation}
where $\mu, \nu, \rho = 4, \ldots, 7$. Thus, the two-forms and five-forms appearing above are trivial in de Rham cohomology. The associated cycles must then vanish in the integer homology of ${\cal N}_3^4$ or generate torsion subgroups. Using the fact that ${\cal N}_3$ and ${\cal N}_3^4$ are compact, closed and orientable, we deduce that the two-cycles $\left(x^2x^3\right)$ and all the five-cycles listed cannot generate torsion and therefore must vanish in the homology. This means that the following M2- and M5-brane wrapping modes must vanish
\begin{equation}
 w^{23} = w^{23456} = w^{23457} = w^{23567} = w^{23467} = w^{14567} = 0 \,.
\end{equation}
Again, these can equivalently be understood as due to the Freed-Witten anomaly in the IIB background obtained by dualising along $\x$ and $\xse$.

Because ${\cal N}_3^4$ is seven-dimensional, we also need to consider Kaluza-Klein monopole wrapping modes. Given that ${\cal N}_3^4$ is orientable, there exists a Kaluza-Klein monopole wrapping mode for each $S^1$ on ${\cal N}_3^4$. For us, it will be important to note that the Nilmanifold ${\cal N}_3$ only has one well-defined $S^1$ which corresponds to the fibre of the principal $\mathrm{U}(1)$-bundle $\pi: {\cal N}_3 \longrightarrow T^2$.

To see that the $S^1$'s on the base do not define well-defined $S^1$'s on ${\cal N}_3$, note that the fibre bundle can be defined symmetrically between the two $S^1$'s on ${\cal N}_3$. This can be seen by making the coordinate redefinition $\x \longrightarrow u = \x + N \y \z$ on the fibre. Then the identifications become
\begin{equation}
 \left( u ,\,\y,\,\z\right) \sim \left( u +1,\,\y,\,\z\right) \sim \left( u + \frac12 N \z,\,\y,\,\z+1\right) \sim \left( u -\frac12 N\z,\,\y+1,\,\z\right) \,.
\end{equation}
This symmetry means that if one of the two $S^1$'s of the $T^2$ is well-defined on ${\cal N}_3$ then both must be well-defined. However, this is impossible since otherwise there would be an inclusion map $\imath: T^2 \longrightarrow {\cal N}_3$ which in fact defines a global section of $\pi: {\cal N}_3 \longrightarrow T^2$. But since this is a non-trivial principal bundle, a global section cannot exist and therefore neither of the $S^1$'s of the base $T^2$ is well-defined on ${\cal N}_3$. This implies that the Kaluza-Klein monopoles cannot wrap the $\y$ or $\z$ cycles. Denoting the Kaluza-Klein monopole wrapping modes by $w^i_{KK}$ where $x^i$ is the local coordinate on the $S^1$ associated with the monopole, we find $w^{2}_{KK} = w^{3}_{KK} = 0$.

We can also see that $w^2_{KK} = w^3_{KK} = 0$ from dualities. Consider applying a U-duality along the three directions $\x$, $\y$, $\w$ of ${\cal N}_3^4$. This maps ${\cal N}_3^4$ to itself while exchanging the M5-brane and Kaluza-Klein monopole wrapping modes
\begin{equation}
 w^{23567} \longleftrightarrow w^2_{KK} \,.
\end{equation}
We have seen that $w^{23567} = 0$ which by duality also requires $w^2_{KK} = 0$. One can make an analogous argument, or appeal to the symmetry of this Nilmanifold, to show that $w^3_{KK} = 0$ as well.

\subsubsection{Wrapping states of ${\cal N}_2^3$} \label{s:N23}
We will construct a five-dimensional locally non-geometric background by dualising ${\cal N}_2^3$. To find missing momenta, we want to know what wrapping states are forbidden on ${\cal N}_2^3$.

Because ${\cal N}_2$ is non-compact, with $\y$ the non-compact coordinate, it is clear that we cannot have any wrapping states involving the $\y$ direction. Therefore we find in particular that the membrane and 5-brane wrapping modes $w^{12} = w^{12345} = 0$ must vanish.

\subsubsection{Wrapping states of ${\cal N}_2^5$} \label{s:N25}
We will dualise ${\cal N}_2^5$ in section \ref{s:E7Chain1} to obtain seven-dimensional locally non-geometric backgrounds. Because of these dualities certain missing wrapping modes of ${\cal N}_2^5$ will become missing momenta. Once again, we note that ${\cal N}_2$ is non-compact with $\y$ the non-compact coordinate. Therefore, there are no states wrapping the $\y$ direction, in particular
\begin{equation}
 w^{12} = w^{12\mu\nu\rho} = w^2_{KK} = 0 \,,
\end{equation}
where $\mu, \nu, \rho = 3, \ldots, 7$ and $w^i_{KK}$ denotes a Kaluza-Klein monopole wrapping mode along the $x^i$ direction.

\subsection{Review of four-dimensional duality chain} \label{s:SL5Chain}
We begin by reviewing the duality chain in four dimensions which leads to the simplest example of a locally non-geometric background of M-theory \cite{Blair:2014zba}. We begin with the twisted torus ${\cal N}_3^1 = {\cal N}_3 \times S^1$ with metric
\begin{equation}
 ds^2 = \left( \dx + N \y \dz \right)^2 + \pdy^2 + \pdz^2 + \pdw^2 \,,
\end{equation}
whose ``geometric flux'' is given by
\begin{equation}
 T_{23}{}^1 = N \,.
\end{equation}

To take an M-theory background to another M-theory background, we need to perform a duality along three directions. To see this, consider the case where the M-theory background consists of a circle fibration over a IIA background. In this case if we take the duality to act along the M-theory circle, as well as two directions of the IIA background, we find a double T-duality relating two different IIA string theory backgrounds.

Applying a duality along the $\y$, $\z$ and $\w$ directions of ${\cal N}_3^1$  takes $\y \longrightarrow \tilde{x}_{34}$ since $\y$ is not an isometry. It also transforms the metric and 3-form to
\begin{equation}
 \begin{split}
  ds^2 &= \left( 1 + N \tilde{x}_{34} \right)^{1/3} \pdx^2 + \left( 1 + N \tilde{x}_{34} \right)^{-2/3} \left[ \pdy^2 + \pdz^2 + \pdw^2 \right] \,, \\
  C_{(3)} &= \frac{N \tilde{x}_{34}}{1 + N^2 \tilde{x}_{34}} \dy \wedge \dz \wedge \dw \,.
 \end{split}
\end{equation}
However, the metric and 3-form are ill-defined as one transverses the dual circle $\tilde{x}_{34} \longrightarrow \tilde{x}_{34} + 1$, where one would have to patch with a U-duality, just as for the type II non-geometry \eqref{eq:QFlux} and \eqref{eq:RFluxT3}. Furthermore, just like in that case, a $\SO{5}$ rotation (which is also not globally well-defined), can be used to change to the non-geometric parameterisation. The new fields are given by \cite{Malek:2012pw}
\begin{equation}
 \begin{split}
  \hat{g}_{ij} &= \left( 1 + V^2 \right)^{-1/3} \left[ \left(1+V^2\right) g_{ij} - V_i V_j \right] \,, \\
  \Omega^{ijk} &= \left(1+V^2\right)^{-1} g^{il} g^{jm} g^{kn} C_{lmn} \,,
 \end{split}
\end{equation}
where $V^i = \frac{1}{3!\sqrt{g}} \epsilon^{ijkl} C_{jkl}$ and $i, j, k = 1 , \ldots, 4$ indices are raised and lowered with $g_{ij}$. Using these formulate we find
\begin{equation}
 \begin{split}
  ds^2 &= \pdx^2 + \pdy^2 + \pdz^2 + \pdw^2 \,, \\
  \Omega^{124} &= - N \tilde{x}_{34} \,. \label{eq:T4NonGeometric}
 \end{split}
\end{equation}
We can also obtain this result by applying the transformation rules \eqref{eq:3DualityRuleTriSimple} and \eqref{eq:3DualityRuleX} discussed in appendix \ref{A:DualityRules}. The background \eqref{eq:T4NonGeometric} has non-vanishing $R$-flux
\begin{equation}
 R^{4,1234} = 4 \hpartial^{4[1} \Omega^{234]} = N \,.
\end{equation}
We summarise this by writing
\begin{equation}
 T_{23}{}^1 \xrightarrow{U_{234}} R^{4,1234} \,,
\end{equation}
where $U_{234}$ denotes the U-duality along $\y$, $\z$ and $\w$.

In section \ref{s:N33} we showed that ${\cal N}_3^1$ does not admit a M2-brane wrapping the $\left(\y\z\right)$-cycle, i.e. $w^{23} = 0$. As discussed in \cite{Gunaydin:2016axc}, this leads to a missing momentum in the locally non-geometric background since the U-duality along $\y,\,\z,\,\w$ exchanges
\begin{equation}
 w^{23} \longrightarrow p_4 \,.
\end{equation}
Therefore, we find $p_4 = 0$ and in \cite{Gunaydin:2016axc} it was argued that this can be written covariantly as
\begin{equation}
 R^{i,jklm} p_i = 0 \,.
\end{equation}

In the following we will show how this generalises to larger compactifications with new $R$-fluxes. We will see that each $R$-flux of mixed symmetry leads to a missing momentum mode.

\subsection{Five-dimensional duality chain} \label{s:SO55Chain}
In section \ref{s:SO55RFlux} we saw that in five-dimensional compactifications, we can have a totally antisymmetric $R$-flux
\begin{equation}
 R^{ijklm} = 20 \hpartial^{[ij} \Omega^{klm]} - 2 \epsilon^{ijklm} \hpartial^z |e| \,.
\end{equation}
We want to consider a duality chain that leads to a locally non-geometric background with this new $R$-flux so that it is not a trivial embedding of the four-dimensional case of \cite{Blair:2014zba,Gunaydin:2016axc}. To this end we begin with the parallelisable but non-unimodular space ${\cal N}_2^3$ with metric
\begin{equation}
 ds^2 = \left( \dx + N \x \dy \right)^2 + \pdy^2 + \pdz^2 + \pdw^2 + \pdxf^2 \,,
\end{equation}
and geometric flux
\begin{equation}
 T_{12}{}^1 = N \,.
\end{equation}

Performing a duality along $\y$, $\z$ and $\w$, and using the U-duality rules \eqref{eq:3DualityRuleTriSimple} and \eqref{eq:3DualityRuleX}, we find that
\begin{equation}
 e^{\bar{1}}{}_2 \longrightarrow \Omega^{134} = N \x \,.
\end{equation}
This is a globally non-geometric, or ``$Q$-flux'', background and we must dualise again, along the directions $\x$, $\y$, $\xf$ to obtain a locally non-geometric background. We find from \eqref{eq:3DualityRuleTriSimple} and \eqref{eq:3DualityRuleX} that this just takes $\x \longrightarrow \tilde{x}_{25}$ while leaving $g_{ij}$ and $\Omega^{ijk}$ unchanged, so that the background is given by
\begin{equation}
 \begin{split}
  ds^2 &= \pdx^2 + \pdy^2 + \pdz^2 + \pdw^2 + \pdxf^2 \,, \\
  \Omega^{134} &= N \tilde{x}_{25} \,.
 \end{split}
\end{equation}
This cannot be obtained in the $T^4$ example of section \ref{s:SL5Chain} because it requires two dualities with only one common direction. This background has non-vanishing $R$-fluxes
\begin{equation}
 \begin{split}
  R^{2,1345} &= - N \,, \\
  R^{5,1234} &= N \,.
 \end{split}
\end{equation}
Note that this implies that $R^{ijklm} = 5 R^{[i,jklm]}$ does not vanish, since
\begin{equation}
 R^{12345} = 2 N \,.
\end{equation}
We summary this duality chain by
\begin{equation}
 T_{12}{}^1 \xrightarrow{U_{234}} Q_1{}^{134} \xrightarrow{U_{125}} R^{2,3145} = - R^{5,1234} \,.
\end{equation}

As we have shown in \ref{s:N23}, the space ${\cal N}_2^3$ is missing the wrapping modes $w^{12} = w^{12345} = 0$. After the dualities along $\y,\,\z,\,\w$ and then $\x,\,\y,\,\xf$, we find from \eqref{eq:3DualityRuleP} that
\begin{equation}
 w^{12} \longrightarrow p_5 \,, \qquad w^{12345} \longrightarrow p_2 \,,
\end{equation}
so that the missing wrapping modes become two missing momenta in the locally non-geometric background. Therefore the conjectured relation of \cite{Gunaydin:2016axc}
\begin{equation}
 R^{i,jklm} p_i = 0 \,,
\end{equation}
holds even in this case.

\subsection{Six-dimensional duality chain} \label{s:E66Chain}
We saw that in $\EG{6}$ there is a new kind of $R$-flux with mixed symmetry given by equation \eqref{eq:E6RFlux}
\begin{equation}
 R^{ij,klmnpq} = 6 \hpartial^{[klmnp} \Omega^{q]ij} +10 \hpartial^{ij} \Omega^{klmnpq} +20 \Omega^{[klm} \hpartial^{|ij|} \Omega^{npq]} \,.
\end{equation}
We will now construct a duality chain which leads to this kind of locally non-geometric background starting with the twisted torus ${\cal N}_3^3$, with metric
\begin{equation}
 ds^2 = \left( \dx + N \y \dz \right)^2 + \pdy^2 + \pdz^2 + \pdw^2 + \pdxf^2 + \pdxsi^2 \,,
\end{equation}
and geometric flux $T_{23}{}^1 = N$.

Because we are considering a six-dimensional compactification, we can either act with U-dualities along three coordinates or all six coordinates. For us it is sufficient to dualise only along three coordinates. We begin with a duality along the three coordinates $\y$, $\z$ and $\w$ to obtain the locally non-geometric $R$-flux as in section \ref{s:SL5Chain}:
\begin{equation}
 \begin{split}
  ds^2 &= \pdx^2 + \pdy^2 + \pdz^2 + \pdw^2 + \pdxf^2 + \pdxsi^2 \,, \\
  \Omega^{124} &= - N \tilde{x}_{34} \,.
 \end{split}
\end{equation}
However, we can now act with a U-duality along $\x$, $\xf$ and $\xsi$ which according to \eqref{eq:3DualityRuleX} takes
\begin{equation}
 \tilde{x}_{34} \longrightarrow \tilde{x}_{13456} \,,
\end{equation}
so that, according to \eqref{eq:3DualityRuleTriSimple}, the trivector becomes
\begin{equation}
 \Omega^{124} = - N \tilde{x}_{13456} \,.
\end{equation}
This is a locally non-geometric background carrying the new $R$-flux
\begin{equation}
 R^{14,123456} = 6 \partial^{[12345} \Omega^{6]14} = N \,.
\end{equation}
We can summarise this duality chain by
\begin{equation}
 T_{23}{}^1 \xrightarrow{U_{234}} R^{4,1234} \xrightarrow{U_{156}} R^{14,123456} \,.
\end{equation}

As we showed in section \ref{s:N33}, the twisted torus ${\cal N}_3^3$ is missing the following M2-brane and M5-brane winding numbers
\begin{equation}
 w^{23} = 0 \,, \qquad w^{23456} = 0 \,.
\end{equation}
After U-duality along $\y$, $\z$, $\w$, followed by U-duality along $\x$, $\xf$ and $\xsi$ we find from \eqref{eq:3DualityRuleP} that the missing winding numbers transform into
\begin{equation}
 w^{23} \longrightarrow p_4 \,, \qquad w^{23456} \longrightarrow p_1 \,,
\end{equation}
and hence we have two missing momenta $p_4$ and $p_1$. This matches
\begin{equation}
 R^{ij,klmnpq} p_i = 0 \,,
\end{equation}
providing a natural generalisation of the conjecture of \cite{Gunaydin:2016axc}.

\subsection{Seven-dimensional duality chain} \label{s:E7Chain}
Here we will construct two duality chains that lead to the new locally non-geometric backgrounds in seven dimensions, for which some of the new $R$-fluxes
\begin{equation}
 \begin{split}
  R^{ij}{}_j &= \hpartial_{jk} \Omega^{ijk} - \frac1{72} \epsilon_{jklmnpq} \hpartial^{ij} \Omega^{klmnpq} - \frac{1}{36} \epsilon_{jklmnpq} \Omega^{klm} \hpartial^{ij} \Omega^{npq} \\
  & \quad - 2 \hpartial^{i} |e| - 10 e_{\bi}{}^{i} \hpartial^j e^{\bi}{}_j \,, \\
  R^i &= \hpartial_{jk} \Omega^{ijk} - 4 e_{\bi}{}^j \hpartial^i e^{\bi}{}_j - 8 e_{\bi}{}^i \hpartial^j e^{\bi}{}_j \,, \\
  R &= \epsilon_{ijklmnp} \hpartial^{i} \Omega^{jklmnp} - 2 \epsilon_{ijklmnp} \Omega^{ijk} \hpartial^m \Omega^{lnp} \,, \\
  R^{ijkl} &= \hpartial^{[i} \Omega^{jkl]} + \frac45 \hpartial_{pq} \Omega^{pqijkl} + \frac25 \Omega^{[ijk} \hpartial_{pq} \Omega^{l]pq} \,,
 \end{split} \label{eq:7dLNG}
\end{equation}
are non-zero.

\subsubsection{Duality chain for $R^{ij}{}_j$ and $R^i$} \label{s:E7Chain1}
To obtain $R^{ij}{}_j \neq 0$ or $R^i \neq 0$ we start with ${\cal N}_2^5$, whose metric can be taken to be
\begin{equation}
 ds^2 = \left( \dx + N \x \dy \right)^2 + \pdy^2 + \pdz^2 + \ldots + \left( \dxse \right)^2 \,,
\end{equation}
and which has non-unimodular geometric flux $T_{12}{}^1 = N$. As in section \ref{s:SO55Chain} we dualise along $\y$, $\z$, $\w$ first and then along $\x$, $\y$ and $\xf$ to obtain the locally non-geometric background
\begin{equation}
 \begin{split}
  ds^2 &= \pdx^2 + \pdy^2 + \pdz^2 + \ldots + \left( \dxse \right)^2 \,, \\
  \Omega^{134} &= N \tilde{x}_{25} \,.
 \end{split}
\end{equation}
This space has $R^{5,1234} = - R^{2,1345} = N$ and is missing the corresponding momentum modes
\begin{equation}
 R^{i,jklm} p_i = 0 \,.
\end{equation}

We now perform a further duality along $\y$, $\xsi$ and $\xse$ which yields the locally non-geometric space
\begin{equation}
 \begin{split}
  ds^2 &= \pdx^2 + \pdy^2 + \pdz^2 + \ldots + \left( \dxse \right)^2 \,, \\
  \Omega^{123467} &= N \tilde{x}_{25} \,. \label{eq:R255}
 \end{split}
\end{equation}
From \eqref{eq:7dLNG} we see that this space has $R^{ij}{}_k$ with the only non-vanishing components given by
\begin{equation}
 R^{25}{}_5 = N \,.
\end{equation}
This is therefore a new kind of locally non-geometric space, not accessible in six dimensions.

Under the above dualities, we also find from \eqref{eq:3DualityRuleP} that the wrapping modes
\begin{equation}
 \begin{split}
  w^{12} \longrightarrow p_5 \,, \qquad w^2_{KK} \longrightarrow p_2 \,,
 \end{split}
\end{equation}
become momenta. As discussed in section \ref{s:N25}, the topology of ${\cal N}_2^5$ implies that both $w^{12} = w^2_{KK} = 0$. Therefore in the locally non-geometric space \eqref{eq:R255} we have
\begin{equation}
 R^{ij}{}_k p_i = 0 \,. \label{eq:RijkMissP}
\end{equation}
However, this could be modified when $R^i \neq 0$ since it could be a linear combination $\tilde{R}^{ij}{}_k = R^{ij}{}_k + \alpha R^{[i} \delta^{j]}_k$, for some $\alpha$, that satisfies
\begin{equation}
 \tilde{R}^{ij}{}_k p_i = 0 \,.
\end{equation}

To check this we act on \eqref{eq:R255} with another duality along $\x$, $\z$ and $\w$ and obtain
\begin{equation}
 \begin{split}
  ds^2 &= \pdx^2 + \pdy^2 + \pdz^2 + \ldots + \left( \dxse \right)^2 \,, \\
  \Omega^{267} &= N \tilde{x}_{12345} \,. \label{eq:R2R266R277}
 \end{split}
\end{equation}
From \eqref{eq:7dLNG} we see that this has non-vanishing $R^i$ and $R^{ij}{}_k$ with non-zero components
\begin{equation}
 R^2 = N \,, \qquad R^{26}{}_6 = R^{27}{}_7 = N \,.
\end{equation}
Therefore we have extended the duality chain of \ref{s:SO55Chain} as
\begin{equation}
 T_{12}{}^{1} \xrightarrow{U_{234}} Q_2{}^{234} \xrightarrow{U_{125}} R^{5,1234} = - R^{2,1345} \xrightarrow{U_{267}} R^{25}{}_5 \xrightarrow{U_{134}} R^2 = R^{26}{}_6 = R^{27}{}_7 \,.
\end{equation}

We see from \eqref{eq:3DualityRuleP} that under these dualities the following wrapping modes of ${\cal N}_2^5$ become momenta:
\begin{equation}
 \begin{split}
  w^2_{KK} &\longrightarrow p_2 \,, \qquad w^{12} \longrightarrow p_5 \,, \qquad w^{12567} \longrightarrow p_1 \,,\\
  w^{12467} &\longrightarrow p_3 \,, \qquad w^{12367} \longrightarrow p_4 \,.
 \end{split}
\end{equation}
However, the topology of ${\cal N}_2^5$ implies that these wrapping modes must vanish and as a result $p_1 = p_2 = p_3 = p_4 = p_5 = 0$ in the locally non-geometric background \eqref{eq:R2R266R277}. Therefore, the momenta satisfy
\begin{equation}
 \left( R^{ij}{}_k - 2 R^{[i} \delta^{j]}_k \right) p_i = 0 \,. \label{eq:RiMissP}
\end{equation}

\subsubsection{Duality chain for $R^{ijkl}$ and $R$} \label{s:E7Chain2}
We begin with the space ${\cal N}_3^4 = {\cal N}_3 \times T^4$ with geometric flux
\begin{equation}
 T_{23}{}^1 = N \,,
\end{equation}
and whose wrapping modes were discussed in section \ref{s:N34}. Performing a U-duality along the six directions $\y$, $\z$, $\w$, $\xf$, $\xsi$, $\xse$ we see from \eqref{eq:6DualityRuleSix} and \eqref{eq:6DualityRuleX} that we obtain the space
\begin{equation}
 \begin{split}
  ds^2 &= \pdx^2 + \pdy^2 + \pdz^2 + \pdw^2 + \pdxf^2 + \pdxsi^2 + \pdxse^2 \,, \\
  \Omega^{124567} &= N \tilde{x}_{34567} \,.
 \end{split}
\end{equation}
This space has non-vanishing locally non-geometric flux
\begin{equation}
 R^{4567} = N \,.
\end{equation}
The duality also exchanges the following M5-wrapping modes for momenta
\begin{equation}
 w^{23567} \longrightarrow p_4 \,, \qquad w^{23467} \longrightarrow p_5 \,, \qquad w^{23457} \longrightarrow p_6 \,, \qquad w^{23456} \longrightarrow p_7 \,,
\end{equation}
according to \eqref{eq:6DualityRuleP}. In section \ref{s:N34} we saw that none of these M5-wrapping states are allowed on ${\cal N}_3^4$ and therefore the momenta $p_4$, $p_5$, $p_6$, $p_7$ are missing in the $R$-flux background. This corresponds exactly to
\begin{equation}
 R^{ijkl} p_l = 0 \,.
\end{equation}
Since $R^{ijkl}$ is really a mixed symmetry tensor $R^{ijkl,mnpqrst}$ with
\begin{equation}
 R^{ijkl} = \frac1{7!} \epsilon_{mnpqrst} R^{ijkl,mnpqrst} \,,
\end{equation}
we see that this confirms our expectations that there are missing momenta for each mixed symmetry $R$-flux.

Note also that, according to \eqref{eq:6DualityRuleP}, the missing M2-brane and KK-monopole wrapping modes become
\begin{equation}
 w^{23} \longrightarrow w^{23} \,, \qquad w^3_{KK} \longrightarrow w^{13} \,, \qquad w^2_{KK} \longrightarrow w^{12} \,,
\end{equation}
so that this non-geometric background has $w^{23} = w^{13} = w^{12} = 0$.

We can now act with another U-duality along the $\x$, $\y$ and $\z$ directions to turn $R^{4567} \longrightarrow R$. In particular, we find that this does not change the non-geometric metric or $\Omega^{ijklmnp}$, but simply changes the dual coordinate $\tilde{x}_{34567} \longrightarrow \tilde{x}_3$, see appendix \ref{A:6Duality}. We find
\begin{equation}
 \begin{split}
  ds^2 &= \pdx^2 + \pdy^2 + \pdz^2 + \pdw^2 + \pdxf^2 + \pdxsi^2 + \pdxse^2 \,, \\
  \Omega^{124567} &= - N \tilde{x}_{3} \,.
 \end{split}
\end{equation}
This space has
\begin{equation}
 R = N \,.
\end{equation}
The entire duality chain is summarised by
\begin{equation}
 T_{23}{}^{1} \xrightarrow{U_{234567}} R^{4567} \xrightarrow{U_{123}} R^{1234567} \,.
\end{equation}

Equation \eqref{eq:3DualityRuleP} also shows that the U-duality takes
\begin{equation}
 w^{23} \longrightarrow p_1 \,, \qquad w^{13} \longrightarrow p_2 \,, \qquad w^{12} \longrightarrow p_3 \,,
\end{equation}
while not acting on $p_4$, $p_5$, $p_6$ and $p_7$. From the previous results we see that all these momentum modes are now missing. We conclude that
\begin{equation}
 R\, p_i = 0 \,,
\end{equation}
so that there are no momentum modes in these locally non-geometric backgrounds with $R \neq 0$.

\section{Conclusions} \label{s:Conclusions}
In this paper we studied locally non-geometric spaces of higher dimensions in M-theory and found a variety of new $R$-fluxes characterising these. Our results can be nicely summarised in seven dimensions where non-geometric backgrounds are parameterised by a trivector $\Omega^{ijk}$ and six-vector $\Omega^{ijklmn}$ transforming as
\begin{equation}
 \begin{split}
  \delta_\xi \Omega^{ijk} &= L_\xi \Omega^{ijk} - 3 \partial^{[ij} \xi^{k]} \,, \\
  \delta_\xi \Omega^{ijklmn} &= L_\xi \Omega^{ijklmn} - \frac1{10} \epsilon^{ijklmnp} \partial_{pq} \xi^q - 6 \Omega^{[ijk} \partial^{lm} \xi^{n]} \,, 
 \end{split}
\end{equation}
under spacetime diffeomorphisms, where
\begin{equation}
 \begin{split}
  L_\xi \Omega^{ijk} &= \xi^l \partial_l \Omega^{ijk} - 3 \Omega^{l[ij} \partial_l \xi^{k]} \,, \\
  L_\xi \Omega^{ijklmn} &= \xi^p \partial_p \Omega^{ijklmn} - 6 \Omega^{p[ijklm} \partial_p \xi^{n]} \,,
 \end{split}
\end{equation}
denotes the covariant transformation under the Lie derivative. As a result, the locally non-geometric $R$-fluxes are given by the following spacetime tensors
\begin{equation}
 \begin{split}
  R^{i,jklm} &= 4 \hpartial^{i[j} \Omega^{klm]} - e_{\bi}{}^{[j} \epsilon^{klm]inpq} \hpartial_{pq} e^{\bi}{}_n \,, \\
  R^{ij}{}_k &= \hpartial_{kl} \Omega^{ijl} - \frac1{72} \epsilon_{klmnpqr} \hpartial^{ij} \Omega^{lmnpqr} - \frac{1}{36} \epsilon_{klmnpqr} \Omega^{lmn} \hpartial^{ij} \Omega^{pqr} \\
  & \quad + 4 e_{\bi}{}^{[i} \hpartial^{j]} e^{\bi}{}_k + 4 \delta^{[i}_k e_{\bi}{}^{j]} \hpartial^l e^{\bi}{}_l \,, \\
  R^i &= \hpartial_{jk} \Omega^{ijk} - 4 e_{\bi}{}^j \hpartial^i e^{\bi}{}_j - 8 e_{\bi}{}^i \hpartial^j e^{\bi}{}_j \,, \\
  R &= \epsilon_{ijklmnp} \hpartial^{i} \Omega^{jklmnp} - 2 \epsilon_{ijklmnp} \Omega^{ijk} \hpartial^m \Omega^{lnp} \,, \\
  R^{ijkl} &= \frac58 \hpartial^{[i} \Omega^{jkl]} + \frac12 \hpartial_{pq} \Omega^{pqijkl} + \frac14 \Omega^{[ijk} \hpartial_{pq} \Omega^{l]pq} \,.
 \end{split}\label{eq:RFluxSummary}
\end{equation}
If we consider a reduction to IIA string theory, only $R^{i,jklm}$ reduces to the usual string $R$-flux \cite{Shelton:2005cf}. The other $R$-fluxes in general involve either dual coordinates involving D-brane wrapping modes or non-geometric fields associated to the R-R sector.

We showed that examples of new locally non-geometric backgrounds, for which these new $R$-fluxes are non-vanishing, can be obtained by dualising product spaces of nil/solvmanifolds with tori. These can be summarised by
\begin{equation}
 \begin{split}
  T_{12}{}^{1} &\xrightarrow{U_{234}} Q_2{}^{234} \xrightarrow{U_{125}} R^{5,1234} = - R^{2,1345} \xrightarrow{U_{267}} R^{25}{}_5 \xrightarrow{U_{134}} R^2 = R^{26}{}_6 = R^{27}{}_7 \,, \\
  T_{23}{}^1 &\xrightarrow{U_{234}} R^{4,1234} \xrightarrow{U_{156}} R^{14}{}_7 \,, \\
  T_{23}{}^{1} &\xrightarrow{U_{234567}} R^{4567} \xrightarrow{U_{123}} R^{1234567} \,.
 \end{split}
\end{equation}
The topology of the initial geometric background meant that certain brane wrapping states were not allowed. After duality this implies that some momenta were missing in the locally non-geometric background, satisfying (in seven dimensions)
\begin{equation}
 \begin{split}
  R^{i,jklm}\, p_i &= 0 \,, \\
  \left( R^{ij}{}_k - 2 R^{[i} \delta^{j]}_k \right) p_i &= 0 \,, \\
  R^{ijkl}\, p_i &= 0 \,, \\
  R\, p_i &= 0 \,, \label{eq:MissingMomentaSummary}
 \end{split}
\end{equation}
generalising the observation in \cite{Gunaydin:2016axc}. Note that in general dimensions $R^i$ and $R^{ijkl}$ become mixed symmetry tensors $R^{i,jklmnpq}$ and $R^{ijkl,mnpqrst}$, respectively. Therefore each mixed symmetry $R$-flux tensor led to missing momenta according to \eqref{eq:MissingMomentaSummary} with the exception of the seven-dimensional singlet flux $R$ for which \emph{all} momenta vanish.

The fact there is a missing momentum mode in four-dimensional locally non-geometric M-theory backgrounds
\begin{equation}
 R^{i,jklm} p_i = 0 \,, \label{eq:4dMissingMomentum}
\end{equation}
means that the phase space of M2-branes is seven-dimensional. This played an important role in the conjecture of \cite{Gunaydin:2016axc} that the phase space algebra of M2-branes in such spaces is isomorphic to the non-associative algebra of imaginary octonions. Furthermore, include the missing momentum mode leads to an eight-dimensional phase space governed by a 3-algebra \cite{Kupriyanov:2017oob} with $\Spin{8}$ structure. Implementing the constraint \eqref{eq:4dMissingMomentum} reduces this 3-algebra to the non-associative M-theory $R$-flux algebra isomorphic to the imaginary octonions. Similarly, the conditions \eqref{eq:MissingMomentaSummary} have implications for the possible non-associative algebras governing the higher-dimensional locally non-geometric backgrounds. We leave the exploration of this issue for further work.

One can also use this formalism to obtain new locally non-geometric fluxes in IIB string theory, using the IIB solution of the EFT section condition \cite{Blair:2013gqa,Hohm:2013vpa,Hohm:2013uia}. This would extend the analysis of \cite{Aldazabal:2010ef} to include also locally non-geometric fluxes. For example, in \cite{Blair:2014zba} it was shown that already in three-dimensional IIB compactifications one has in addition to the usual $R^{ijk}$ flux, a mixed symmetry flux $R^{i,jk}$ with $R^{[i,jk]} = 0$. It is straightforward to dualise the Nilmanifold ${\cal N}_3$ in a way to obtain a locally non-geometric background with $R^{1,13} \neq 0$. The homology of ${\cal N}_3$ then shows that
\begin{equation}
 R^{i,jk} p_i = 0 \,,
\end{equation}
analogous to \eqref{eq:MissingMomentaSummary}. We leave an investigation of other IIB $R$-fluxes, associated missing momenta and non-associative algebras to future work.

Another interesting question is whether the relations identifying the missing momenta \eqref{eq:MissingMomentaSummary} can be derived directly using a notion of (co-)homology in exceptional field theory. For example, one would expect that the cohomology of an appropriate lift of the usual cochain complex of differential forms to appropriate tensors in exceptional field theory would be able to capture \eqref{eq:MissingMomentaSummary} since it would know of the topology of the twisted tori that are dual to the locally non-geometric backgrounds. Furthermore, as we mentioned in the introduction, the lack of momenta can also in some cases be related to the Freed-Witten anomaly of a dual IIB background. Therefore, this exceptional cohomology would also provide a simple geometric notion within exceptional field theory of these anomaly cancellation conditions.

As has already been argued in \cite{Cederwall:2013naa,Aldazabal:2013via,Hohm:2015xna,Wang:2015hca,Bosque:2016fpi,Malek:2016bpu,Malek:2016vsh,Malek:2017njj}, the tensor hierarchy of exceptional field theory defines a chain complex and its tensors can be thought of as the exceptional generalisation of differential forms. To capture the missing momenta, it seems likely that one should consider generalised vector fields $V$, since these contain two-forms, five-forms, etc., which are ``closed'' but not exact. Exactness is easy to define using the tensor hierarchy, so that an exact generalised vector field satisfies $V = d \chi$ for $\chi \in \Gamma\left({\cal R}_2\right)$, where ${\cal R}_2$ is an appropriate vector bundle from the tensor hierarchy. However, the chain complex of the tensor hierarchy usually is taken to end at precisely the generalised vector fields because there is no non-trivial nilpotent derivative that one can define on generalised vector fields without requiring more structure. It is also worth emphasising that our results regarding missing momenta should also hold for backgrounds which are not generalised parallelisable. For example, a non-trivial $\mathrm{U}(1)$-fibration would then generalise the spaces with geometric flux. In the total space, the 1st Chern class is trivial and hence the homology is reduced, just like we found for the twisted torus. The dual locally non-geometric space will thus similarly have missing momenta. We leave these interesting problems for future work.

Finally, another interesting question would be to understand what branes source the new $R$-fluxes \eqref{eq:RFluxSummary}, generalising the known NS-NS ``R-brane'' \cite{Hassler:2013wsa,Bakhmatov:2016kfn}. To answer this one could consider dualising a Kaluza-Klein monopole, which is a 1/2-BPS analogue of the Heisenberg Nilmanifold, in a similar way to the duality chains considered here. An efficient description of the duality chain may come by using the half-maximal structures in EFT \cite{Malek:2016bpu,Malek:2016vsh,Malek:2017njj}. Related to this, it would be interesting to try and find honest locally non-geometric M-theory backgrounds. For example, one may want to consider dualities of the ${\cal N}=2$ background given in \cite{Kachru:2002sk}.

\section*{Acknowledgements}
The work of DL and EM is supported by the ERC Advanced Grant No.~320045 ``Strings and Gravity''.

\appendix

\section{Details of decomposition of $\EG{d}$ EFT to $\SL{d} \times \mathbb{R}^+$} \label{A:EFT}

\subsection{$\SO{5,5}$ EFT} \label{A:SO55}

The generalised Lie derivative acting on the generalised vielbein $E_{\bar{M}}{}^M$ is given by \cite{Coimbra:2012af,Berman:2012vc}
\begin{equation}
 \begin{split}
 \gL_V E_{\bar{M}}{}^M &= V^N \partial_N E_{\bar{M}}{}^M + \left(\mathbb{P}_{adj}\right)^{M}{}_{N}{}^{P}{}_{Q} E_{\bar{M}}{}^N \partial_P V^Q \\
 &= V^N \partial_N E_{\bar{M}}{}^M - E_{\bar{M}}{}^N \partial_N V^M + \frac12 \left(\gamma^I\right)^{MP} \left(\gamma_I\right)_{NQ} E_{\bar{M}}{}^N \partial_P V^Q - \frac14 E_{\bar{M}}{}^M \partial_N V^N \,, \label{eq:SO55Lie}
\end{split}
\end{equation}
where $M, N = 1, \ldots, 16$ labels the spinor and $I = 1, \ldots, 10$ the fundamental representation of $\SO{5,5}$, $\left(\mathbb{P}_{adj}\right)^{M}{}_{N}{}^P{}_{Q}$ is the projector of the tensor product $\mbf{16} \otimes \obf{16}$ onto the adjoint. Here $\left(\gamma_I\right)^{MN}$ are $\SO{5,5}$ $\gamma$-matrices, satisfying
\begin{equation}
 \left(\gamma_{(I}\right)^{MP} \left(\gamma_{J)}\right)_{NP} = \eta_{IJ} \delta^M_N \,.
\end{equation}
where $\eta_{IJ}$ is the $\SO{5,5}$ invariant metric. The final term with coefficient of $-\frac14$ in \eqref{eq:SO55Lie} ensures that
\begin{equation}
 E^{\bar{M}}{}_M \gL_V E_{\bar{M}}{}^M = \gL_V |E| = 0 \,,
\end{equation}
so that we can, and will, take $E_{\bar{M}}{}^M$ to have unit determinant. The section condition is
\begin{equation}
 \left(\gamma_I\right)^{MN} \partial_M \otimes \partial_N = \left(\gamma_I\right)^{MN} \partial_M \partial_N = 0 \,,
\end{equation}
where we take the derivatives to act on two different fields or on the same field as a double derivative.

Decomposing $\SO{5,5} \longrightarrow \SL{5} \times \mathbb{R}^+$, vectors in the $\mbf{16}$ and $\mbf{10}$ of $\SO{5,5}$ decompose as follows
\begin{equation}
 V^M = \left( V^i ,\, V_{ij},\, V_z \right) \,, \qquad V^I = \left( V^i,\, V_i \right) \,.
\end{equation}
In the $\SL{5} \times \mathbb{R}^+$ basis, the $\gamma$-matrices have the following non-zero components.
\begin{equation}
 \begin{split}
  \left(\gamma_I\right)^{MN}: \quad &\left(\gamma^i\right)^j{}_{kl} = 2 \delta^{ij}_{kl} \,, \qquad \left(\gamma_i\right)^j{}_z = \sqrt{2} \delta_i^j \,, \qquad \left(\gamma_i\right)_{jk\,lm} = \frac1{\sqrt{2}} \epsilon_{ijklm} \,, \\
  \left(\gamma_I\right)_{MN}: \quad & \left(\gamma_i\right)_j{}^{kl} = 2 \delta^{kl}_{ij} \,, \qquad \left(\gamma^i\right)_j{}^z = \sqrt{2} \delta^i_j \,, \qquad \left(\gamma^i\right)^{jk\,lm} = \frac1{\sqrt{2}} \epsilon^{ijklm} \,.
 \end{split} \label{eq:SO55Gammas}
\end{equation}
Note that the $\gamma$-matrices are symmetric in the spinor indices, i.e. $\left(\gamma_I\right)^{MN} = \left(\gamma_I\right)^{NM}$. From \eqref{eq:SO55Gammas}, one obtains the section condition as given in \eqref{eq:SO55SectionCondition}. Furthermore, using \eqref{eq:SO55Gammas} and \eqref{eq:SO55Lie} one finds that the generalised Lie derivative corresponding to a spacetime diffeomorphisms $V^M=\left(\xi^i,0,0\right)$ acts on the vielbein as
\begin{equation}
 \begin{split}
  \gL_{\xi} E_{\bar{M}}{}^i &= \xi^j\partial_j E_{\bar{M}}{}^i - E_{\bar{M}}{}^M \partial_M \xi^i -2 E_{\bar{M}\,jk} \partial^{ji} \xi^k +  E_{\bar{M}\,z} \partial^z \xi^i - \frac{1}{4} E_{\bar{M}}{}^i \partial_j \xi^j\,, \\
  \gL_{\xi} E_{\bar{M}\,ij} &= \xi^k \partial_k E_{\bar{M}\,ij} - 2 E_{\bar{M}k[i}\partial_{j]} \xi^k + \frac{1}{2} \epsilon_{ijklm} E_{\bar{M}\,z} \partial^{kl} \xi^m - \frac{1}{4} E_{\bar{M}\,ij} \partial_k \xi^k \,, \\
  \gL_{\xi} E_{\bar{M}\,z} &= \xi^i \partial_i E_{\bar{M}\,z} + \frac{3}{4} E_{\bar{M}\,z} \partial_k \xi^k \,.
 \end{split} \label{eq:SO55LieVielbein}
\end{equation}
From this one can easily recover the transformation law for $\Omega^{ijk}$ given in equation \eqref{eq:SO55OmegaTransform}.

\subsection{$\EG{6}$ EFT} \label{A:E6}

The $\EG{6}$ generalised Lie derivative of the generalised vielbein $E_{\bar{M}}{}^M$ is given by \cite{Coimbra:2012af,Berman:2012vc,Hohm:2013vpa}
\begin{equation}
 \begin{split}
  \gL_V E_{\bar{M}}{}^M &= V^N \partial_N E_{\bar{M}}{}^M + \left(\mathbb{P}_{adj}\right)^{M}{}_{N}{}^{P}{}_{Q} V^N \partial_P E_{\bar{M}}{}^Q \\
  &= V^N \partial_N E_{\bar{M}}{}^M - E_{\bar{M}}{}^N \partial_N V^M + 10 d^{MNP} d_{PQR} E_{\bar{M}}{}^Q \partial_N V^R - \frac13 E_{\bar{M}}{}^M \partial_N V^N \,, \label{eq:E6Lie}
 \end{split}
\end{equation}
where $M, N = 1, \ldots, 27$ label the fundamental representation of $\EG{6}$, $d_{MNP}$ and $d^{MNP}$ are the symmetric cubic $\EG{6}$ invariant tensors, $\left(\mathbb{P}_{adj}\right)^{M}{}_{N}{}^{P}{}_{Q}$ is the projector onto the adjoint. The cubic invariants are normalised as
\begin{equation}
 d^{MPQ} d_{NPQ} = \delta^M_N \,,
\end{equation}
and the section condition is given by
\begin{equation}
 d^{MNP} \partial_N \otimes \partial_P = d^{MNP} \partial_N \partial_P = 0 \,.
\end{equation}
The final term with coefficient $-\frac13$ in the second line of \eqref{eq:E6Lie} ensures that
\begin{equation}
 E^{\bar{M}}{}_M \gL_U E_{\bar{M}}{}^M = \gL_U |E| = 0 \,,
\end{equation}
so that we can, and will, take $E_{\bar{M}}{}^M$ to have unit determinant.

Decomposing $\EG{6} \longrightarrow \SL{6} \times \SL{2} \longrightarrow \SL{6} \times \mathbb{R}^+$, we have that a vector in the fundamental of $\EG{6}$ becomes
\begin{equation}
 V^M = \left( V^i ,\, V_{ij} ,\, V_{ijklm} \right) \,,
\end{equation}
while we take the coordinate derivatives to be
\begin{equation}
 \partial_M = \left( \partial_i,\, \partial^{ij},\, - \frac{1}{\sqrt{5!}} \partial^{ijklm} \right) \,.
\end{equation}
The only non-zero components of cubic invariants are given by
\begin{equation}
 \begin{split}
  d^{MNP}: \quad & d^{i}{}_{jk\,lmnpq} = \frac{1}{10\sqrt{6}} \delta^i_{[j} \epsilon_{k]lmnpq} \,, \qquad d_{ij\,kl\,mn} = \frac1{4\sqrt{5}} \epsilon_{ijklm} \,, \\
  d_{MNP}: \quad & d_i{}^{jk\,lmnpq} = \frac1{10\sqrt{6}} \delta_i^{[j} \epsilon^{k]lmnpq} \,, \qquad d^{ij\,kl\,mn} = \frac{1}{4\sqrt{5}} \epsilon^{ijklmn} \,.
 \end{split} \label{eq:dtensors}
\end{equation}
Using this decomposition, one obtains the section condition as given in \eqref{eq:E6SectionCondition}. Using \eqref{eq:dtensors} and \eqref{eq:E6Lie}, one finds that the generalised Lie derivative corresponding to a spacetime diffeomorphism $V^M = \left( \xi^i,\, 0,\, 0 \right)$ acts on the generalised vielbein as
\begin{equation}
 \begin{split}
  \gL_{\xi} E_{\bar{M}}{}^i &= \xi^j \partial_j E_{\bar{M}}{}^i - E_{\bar{M}}{}^j \partial_j \xi^i - \frac{6}{5!} E_{\bar{M}\,j_1 j_2 j_3 j_4 j_5} \partial^{[ i j_1 j_2 j_3 j_4} \xi^{j_5]}  - 3 E_{\bar{M}\,jk} \partial^{[jk} \xi^{i]} - \frac13 E_{\bar{M}}{}^i \partial_j \xi^j \,, \\
  \gL_{\xi} E_{\bar{M}\,ij} &= \xi^k \partial_k E_{\bar{M}\,ij} + 2 E_{\bar{M}\,k[j} \partial_{i]} \xi^k + E_{\bar{M}\,ik} \partial_j \xi^k - \sqrt{30} E_{\bar{M}\,ijklm} \partial^{kl} \xi^m - \frac13 E_{\bar{M}\,ij} \partial_k \xi^k \,, \\
  \gL_{\xi} E_{\bar{M}\,i_1 i_2 i_3 i_4 i_5} &= \xi^j \partial_j E_{\bar{M}\,i_1i_2i_3i_4i_5} + 5 E_{\bar{M}\,j[i_1 i_2 i_3 i_4}  \partial_{i_5]} \xi^j - \frac{1}{3} E_{\bar{M}\,i_1 i_2 i_3 i_4 i_5} \partial_j \xi^j \,. \label{eq:E6LieVielbein}
 \end{split}
\end{equation}

\subsection{$\EG{7}$ EFT} \label{A:E7}

Finally, the $\EG{7}$ generalised Lie derivative of the generalised vielbein $E_{\bar{M}}{}^M$ is given by
\begin{equation}
 \begin{split}
  \gL_V E_{\bar{M}}{}^M &= V^N \partial_N E_{\bar{M}}{}^M - 12 \left(\mathbb{P}_{adj}\right)^M{}_N{}^P{}_Q E_{\bar{M}}{}^N \partial_P V^Q \\
  &= V^N \partial_N E_{\bar{M}}{}^M - E_{\bar{M}}{}^N \partial_N V^M - 12 \left(t_\alpha\right)^{MP} \left(t^\alpha\right)_{NQ} E_{\bar{M}}{}^N \partial_P V^Q \\
  & \quad + \frac12 \Omega^{MP} \Omega_{NQ} E_{\bar{M}}{}^N \partial_P V^Q - \frac12 E_{\bar{M}}{}^M \partial_N V^N \,,
 \end{split} \label{eq:E7Lie}
\end{equation}
where $M, N = 1, \ldots, 56$ label the fundamental and $\alpha, \beta = 1, \ldots, 133$ label the adjoint representations of $\EG{7}$.
\begin{equation}
 \begin{split}
  \left(\mathbb{P}_{adj}\right)^M{}_N{}^P{}_Q &\equiv \left(t_\alpha\right)^M{}_N \left(t^\alpha\right)^P{}_Q \\
  &= \frac1{24} \delta^M_N \delta^P_Q + \frac1{12} \delta^M_Q \delta^P_Q + \left(t_\alpha\right)^{MP} \left(t^\alpha\right)_{NQ} - \frac1{24} \Omega^{MP} \Omega_{NQ} \,,
 \end{split}
\end{equation}
is the projector of the tensor product $\mbf{56} \otimes \mbf{56}$ onto the adjoint, with $\left(t_\alpha\right)^{MN}$ the generators of $\EG{7}$ in the fundamental representations. $M, N, \ldots$ indices are raised and lowered with the symplectic invariant $\Omega_{MN}$ of $\mathrm{Sp}(56) \supset \EG{7}$ according to a north-west south-east convention, i.e.
\begin{equation}
 V^M = \Omega^{MN} V_N \,, \qquad V_M = V^N \Omega_{MN} \,,
\end{equation}
with
\begin{equation}
 \Omega^{MP} \Omega_{NP} = \delta^M_N \,,
\end{equation}
while the $\EG{7}$ adjoint indices $\alpha, \beta$ are raised/lowered with the Killing metric
\begin{equation}
 \kappa_{\alpha\beta} = \left(t_\alpha\right)^M{}_N \left(t_\beta\right)^N{}_M \,.
\end{equation}
The section condition is given by \cite{Berman:2012vc,Hohm:2013uia}
\begin{equation}
 \left(t_\alpha\right)^{MN} \partial_M \otimes \partial_N = \left(t_\alpha\right)^{MN} \partial_M \partial_N = 0 \,, \qquad \Omega^{MN} \partial_M \otimes \partial_N = 0 \,.
\end{equation}

Under the decomposition $\EG{7} \longrightarrow \SL{8}$, we have
\begin{equation}
 \mbf{56} \longrightarrow \mbf{28} \oplus \obf{28} \,,
\end{equation}
so that a vector $V^M$ can be written as
\begin{equation}
 V^M = \left( V^{IJ} ,\, V_{IJ} \right) \,,
\end{equation}
where $I, J = 1, \ldots, 8$ labels the fundamental of $\SL{8}$ and $V^{IJ} = V^{[IJ]}$ and $V_{IJ} = V_{[IJ]}$ transform in the $\mbf{28}$ and $\obf{28}$ of $\SL{8}$. We can further decompose $\SL{8} \longrightarrow \SL{7} \times \mathbb{R}^+$ so that
\begin{equation} 
 \begin{split}
  V^{IJ} &= \left( V^{i},\, V^{ij} = \frac1{5!} \epsilon^{ijklmnp} V_{klmnp} \right) \,, \\
  V_{IJ} &= \left( V_i,\, V_{ij} \right) \,,
 \end{split}
\end{equation}
and the coordinate derivatives become
\begin{equation}
 \partial_M = \left( \partial_i,\, \partial^{ij} ,\, - \frac12 \partial_{ij} ,\, \partial^i \right) \,.
\end{equation}
Similarly, the adjoint representation of $\EG{7}$ decomposes under $\SL{8}$ as
\begin{equation}
 \mbf{133} \longrightarrow \mbf{63} \oplus \mbf{70} \,.
\end{equation}
In the $\SL{8}$ basis, the only non-zero components of the generators $\left(t_\alpha\right)_{M}{}^{N}$ are
\begin{equation}
 \begin{split} 
  \left(t_{I}{}^{J}\right)_{KL}{}^{MN} & = - \frac{1}{\sqrt{3}} \delta^{J}_{[K} \delta_{L]I}^{MN} - \frac{1}{8\sqrt{3}} \delta_{I}^{J} \delta_{KL}^{MN} = - \left(t_{I}^{J}\right)^{MN}{}_{KL}\,, \\
  \left(t_{IJKL}\right)_{MNPQ} &= \frac{1}{24\sqrt{2}} \epsilon_{IJKLMNPQ},, \qquad \left(t_{IJKL}\right)^{MNPQ} = \frac{1}{\sqrt{2}} \delta_{IJKL}^{MNPQ}\,.
 \end{split}
\end{equation}
The only non-zero components of the symplectic invariant $\Omega_{MN}$ are given by
\begin{equation}
 \Omega^{IJ}{}_{KL} = \delta^{IJ}_{KL} = - \Omega_{KL}{}^{IJ} \,.
\end{equation}

Using this decomposition, the section condition reduces to that given in \eqref{eq:E7SectionCondition}. Furthermore, if we act with the generalised Lie derivative \eqref{eq:E7Lie} corresponding to a spacetime diffeomorphism $V^M = \left( V^{i8} = \frac12 \xi^i ,\, 0,\, 0,\, 0 \right)$ on the generalised vielbein we find
\begin{equation}
 \begin{split}
  \gL_{\xi} E_{\bar{M}}{}^{i} &= \xi^{j} \partial_{j} E_{\bar{M}}{}^{i} - E_{\bar{M}}{}^{j} \partial_{j} \xi^{i} - 3 E_{\bar{M}\,jk} \partial^{[jk} \xi^{i]} - \frac{1}{2} E_{\bar{M}}{}^{i} \partial_{k} \xi^{k} - E_{\bar{M}}{}^{ij} \partial_{jk} \xi^{k} \,, \\
  \gL_{\xi} E_{\bar{M}\,ij} &= \xi^{k} \partial_{k} E_{\bar{M}\,ij} +  2 E_{\bar{M}\,k[j} \partial_{i]} \xi^{k} - \frac12 E_{\bar{M}\,[j} \partial_{i]k} \xi^{k} - \frac{1}{4} \epsilon_{ijklmnp} E_{\bar{M}}{}^{kl} \partial^{mn} \xi^{p} \\
  & \quad - \frac{1}{2} E_{\bar{M}\,ij} \partial_{k} \xi^{k} \,, \\
  \gL_{\xi} E_{\bar{M}}{}^{ij} &= \xi^{k} \partial_{k} E_{\bar{M}}{}^{ij} - 2 E_{\bar{M}}{}^{k[j} \partial_{k} \xi^{i]} - \frac{3}{2} E_{\bar{M}\,k} \partial^{[ij} \xi^{k]} + \frac{1}{2} E_{\bar{M}}{}^{ij} \partial_{k} \xi^{k} \,, \\
  \gL_{\xi} E_{\bar{M}\,i} &= \xi^{j} \partial_{j} E_{\bar{M}\,i} +  E_{\bar{M}\,j} \partial_{i} \xi^{j} + \frac{1}{2} E_{\bar{M}\,i} \partial_{j} \xi^{j} \,.
 \end{split} \label{eq:E7LieVielbein}
\end{equation}

\section{U-duality rules} \label{A:DualityRules}
In the following we will be acting with U-dualities on twisted tori and backgrounds with non-geometric trivector $\Omega^{ijk}$ or even six-vector $\Omega^{ijklmn}$ to generate new locally non-geometric backgrounds in M-theory. The U-dualities must be taken along three or six directions in order to map M-theory backgrounds into one another. Before considering explicit examples, let us introduce an efficient procedure for finding the U-dual backgrounds. We will use the fact that our backgrounds are parallelisable without 4-form or 7-form flux (although these can easily be included), and act with appropriate $\EG{d}$ matrices representing the U-dualities on the generalised vielbein.

For a U-duality along three directions define the totally antisymmetric combinations
\begin{equation}
 \omega_{ijk} = \omega^{ijk} = \pm 1 \,,
\end{equation}
which only have non-vanishing components along the three directions being dualised. For example, for a U-duality along $\x$, $\y$, $\z$ we would have $\omega^{123} = \omega_{123} = 1$. Similarly, for a U-duality along six directions, define the totally antisymmetric combinations
\begin{equation}
 \omega_{ijklmn} = \omega^{ijklmn} = \pm 1 \,,
\end{equation}
whose only non-vanishing components are along the six directions being dualised. The $\EG{d}$ matrices realising the U-duality along three directions are given by \cite{Malek:2012pw}
\begin{equation}
 U = E_{\Omega} E_{C} E_{\Omega} \,,
\end{equation}
where $E_{\Omega}$ and $E_C$ to be the generalised vielbein with non-zero trivector $\Omega$ and three-form
\begin{equation}
 \Omega^{ijk} = - \omega^{ijk} \,, \qquad C_{ijk} = \omega_{ijk} \,.
\end{equation}

Similarly, we take the U-duality along six directions to be realised by the $\EG{d}$ matrix
\begin{equation}
 U = E_{\Omega} E_C E_{\Omega} \,,
\end{equation}
where now $E_{\Omega}$ and $E_C$ to be the generalised vielbein with non-zero six-vector and six-form
\begin{equation}
 \Omega^{ijklmn} = - \omega^{ijklmn} \,, \qquad C_{ijklmn} = \omega_{ijklmn} \,.
\end{equation}

Using this we find the following transformation rules, which we will give for seven-dimensional spaces. This straightforwardly also contains all lower-dimensional cases.

\subsection{U-duality along three directions} \label{A:3Duality}
Consider a parallelisable background such as the ${\cal N}_3^k$ and ${\cal N}_2^k$ spaces in sections \ref{s:N31} -- \ref{s:N25} with metric
\begin{equation}
 ds^2 = R_1^2 \left(e^{\bar{1}}\right)^2 + R_2^2 \left(e^{\bar{2}}\right)^2 + \ldots + R_7^2 \left(e^{\bar{7}}\right)^2 \,,
\end{equation}
where $e^{\bi} = e^{\bi}{}_i dx^i$ are globally well-defined 1-forms.

In this paper we are considering geometric backgrounds that have triangular vielbeine
\begin{equation}
 e^{\bi}{}_j = \delta^{\bi}{}_j + N^{\bi}{}_j \,, \qquad e^i{}_{\bj} = \delta^i{}_{\bj} - N^i{}_{\bj} \,.
\end{equation}
Acting with three dualities on such a space yields a background with trivector and 3-form given by
\begin{equation}
 N^{\bi}{}_j \longrightarrow \left\{ \begin{array}{rl} \Omega'^{ijk} &= 3 N^{[i}{}_{\bl} \, \omega^{jk]l} \,, \\
 C'_{ijk} &= - 3 N^{\bl}{}_{[i} \omega_{jk]\bl} \,, \\
 N'^{\bi}{}_j &= N^{\bi}{}_j + \frac32 N^{[i}{}_{\bm} \omega^{kl]\bm} \omega_{jkl} - \frac32 N^{\bm}{}_{[j} \omega_{kl]\bm} \omega^{\bi kl} \,.
 \end{array} \right. \label{eq:3DualityRuleTriSimple}
\end{equation}
The new coordinates are given by
\begin{equation}
 \begin{split}
  x'^i &= \frac12 \omega^{ijk} \tilde{x}_{jk} \,, \\
  \tilde{x}'_{ij} &= \omega_{ijk} x^k + \frac1{3!} \omega^{klm} \tilde{x}_{ijklm} \,, \\
  \tilde{x}'_{ijklm} &= \omega_{[ijk} \tilde{x}_{lm]} + \frac12 \omega^{npq} \epsilon_{ijklmnp} \tilde{x}_{q} \,, \\
  \tilde{x}'_i &= \frac1{5!2} \omega_{ijk} \epsilon^{jklmnpq} \tilde{x}_{lmnpq} \,. \label{eq:3DualityRuleX}
 \end{split}
\end{equation}
In this paper we take $R_i = 1$. However, for completeness' sake we will give the radii transform laws here too. If we label by $u, v = 1, \ldots, 3$ as the directions along which the duality is taken and $\alpha, \beta = 5, \ldots, 6$ as the directions along which it is not, then the new radii are
\begin{equation}
 R_u \longrightarrow R'_u = R_u \left( R_1\,R_2\,R_3 \right)^{-2/3} \,, \qquad R_\alpha \longrightarrow R'_\alpha = R_\alpha \left( R_1\,R_2\,R_3 \right)^{1/3} \,. \label{eq:3DualityRuleR}
\end{equation}
We see that if the trivector and 3-form in \eqref{eq:3DualityRuleTriSimple} vanish then the twisted torus is left unchanged by the duality, up to the transformations of the coordinates \eqref{eq:3DualityRuleX} and radii \eqref{eq:3DualityRuleR}.

Acting with another duality along three directions on a trivector background (with flat metric) can yield a six-vector or a 3-form
\begin{equation}
 \Omega^{ijk} \longrightarrow \left\{ \begin{array}{c} \Omega'^{ijklmn} = 20 \Omega^{[ijk}\, \omega^{lmn]} \,, \\
 C'_{ijk} = \left( \frac1{3!} \omega_{lmn} \Omega^{lmn} \right) \omega_{ijk} \,,
 \end{array} \right.
\end{equation}
where again if both of these vanish the trivector is left unchanged, up to the coordinate transformation \eqref{eq:3DualityRuleX} and transformation of radii \eqref{eq:3DualityRuleR}.

Momenta and wrapping modes are also exchanged according to the following rules
\begin{equation}
 \begin{split}
  p'_i &= \frac12 \omega_{ijk} w^{jk} \,, \\
  w'^{ij} &= \omega^{ijk} p_k + \frac1{3!} \omega_{klm} w^{ijklm} \,, \\
  w'^{ijklm} &= \omega^{[ijk} w^{lm]} + \frac12 \omega_{npq} \epsilon^{ijklmnp} w^{q} \,, \\
  w'^i &= \frac1{5!2} \omega^{ijk} \epsilon_{ijklmnpq} w^{lmnpq} \,,
 \end{split} \label{eq:3DualityRuleP}
\end{equation}
which are completely analogous to \eqref{eq:3DualityRuleX}.

\subsection{U-duality along six directions} \label{A:6Duality}
If we act with a U-duality on a twisted torus with triangular vielbein
\begin{equation}
 e^{\bi}{}_j = \delta^{\bi}{}_j + N^{\bi}{}_j \,, \qquad e^i{}_{\bj} = \delta^i{}_{\bj} - N^i{}_{\bj} \,,
\end{equation}
along six directions we find
\begin{equation}
 N^{\bar{i}}{}_j \longrightarrow \left\{ \begin{array}{rl} \Omega'^{ijklmn} &= 6 N^{[i}{}_{\bp} \, \omega^{jklmn]\bp} \,, \\
 C'_{ijk} &= - 6 N^{\bp}{}_{[i} \omega_{jklmn]\bp} \,, \\
 N'^{\bi}{}_j &= N^{\bi}{}_j + \frac{6}{5!} N^{[i}{}_{\bar{q}} \omega^{klmnp]\bar{q}} \omega_{jklmnp} - \frac{6}{5!} N^{\bar{q}}{}_{[j} \omega_{klmnp]\bar{q}} \omega^{\bi klmnp} \,.
 \end{array} \right. \label{eq:6DualityRuleSix}
\end{equation}
with new coordinates
\begin{equation}
 \begin{split}
  x'^i &= \frac1{5!}\, \omega^{ijklmn} \tilde{x}_{jklmn} \,, \\
  \tilde{x}'_{ij} &= \frac1{5!}\, \omega^{klmnpq} \epsilon_{ijklmnp} \tilde{x}_q \,, \\
  \tilde{x}'_{ijklm} &= \omega_{ijklmn} x^n \,, \\
  \tilde{x}'_i &= \frac1{5!}\, \omega_{ijklmn} \epsilon^{jklmnpq} \tilde{x}_{pq} \,. \label{eq:6DualityRuleX}
 \end{split}
\end{equation}
Similarly, the momenta and windings are exchanged as
\begin{equation}
 \begin{split}
  p'_i &= \frac1{5!}\, \omega_{ijklmn} w^{jklmn} \,, \\
  w'^{ij} &= \frac1{5!}\, \omega_{klmnpq} \epsilon^{ijklmnp} w^q \,, \\
  w'^{ijklm} &= \omega^{ijklmn} p_n \,, \\
  w'^i &= \frac1{5!}\, \omega^{ijklmn} \epsilon_{jklmnpq} w^{pq} \,.
 \end{split} \label{eq:6DualityRuleP}
\end{equation}
If we label by $u, v = 1, \ldots, 6$ as the directions along which the duality is taken, then the new radii are
\begin{equation}
 \begin{split}
  R_u &\longrightarrow R'_u = R_u \left( R_1\,R_2\,R_3\,R_4\,R_5\,R_6 \right)^{-1/3} \,, \\
  R_7 &\longrightarrow R'_7 = R_7 \left( R_1\,R_2\,R_3\,R_4\,R_5\,R_6 \right)^{2/3} \,.
 \end{split} \label{eq:6DualityRuleR}
\end{equation}

If the 6-vector and 6-form in \eqref{eq:6DualityRuleSix} vanish, then only the coordinates of the twisted torus transform. Furthermore, acting with a duality along six directions of a trivector background would in general generate a 3-form
\begin{equation}
 C'_{ijk} = \frac1{3!}\, \omega_{ijklmn} \Omega^{lmn} \,. \label{eq:SixDualityTri}
\end{equation}
Where this vanishes, the transformation only acts on the coordinates according to \eqref{eq:6DualityRuleX}, leaving $\Omega^{ijk}$ unchanged otherwise.

\providecommand{\href}[2]{#2}\begingroup\raggedright\endgroup

\end{document}